\documentclass[12pt]{iopart}
\expandafter\let\csname equation*\endcsname\relax
\expandafter\let\csname endequation*\endcsname\relax

\usepackage{amsfonts}
\usepackage{amsmath}
\usepackage{amssymb}
\usepackage{bm}
\usepackage{dcolumn}
\usepackage{graphicx}
\usepackage{graphics}
\usepackage[T1]{fontenc}
\usepackage[utf8]{inputenc}
\usepackage{latexsym}
\usepackage{rotating}
\usepackage{cite}
\usepackage[perpage,marginal]{footmisc}
\usepackage[colorlinks=true]{hyperref}
\usepackage[all]{hypcap}
\usepackage{xspace} % Sensible space treatment at end of simple macros
\usepackage[usenames]{color}
\usepackage{mathrsfs}
\usepackage{microtype}
\graphicspath{{fig/}}

\makeatletter
\renewcommand\footnoterule{%
  \kern-3\p@
  \hrule\@width2.5cm
  \kern2.6\p@}
\makeatother

\DeclareMathOperator\arctanh{arctanh}

\definecolor {darkgreen}{rgb}{0.2,0.7,0.2}

\newcommand\be{\begin{equation}}
\newcommand\ba{\begin{eqnarray}}
\newcommand\ee{\end{equation}}
\newcommand\ea{\end{eqnarray}}

\newcommand\bw{\begin{widetext}}
\newcommand\ew{\end{widetext}}

\newcommand{\nn}{\nonumber}

\newcommand{\Kerr}{{\mbox{\tiny Kerr}}}

\newcommand{\GR}{{\mbox{\tiny GR}}}

\newcommand{\CS}{{\mbox{\tiny CS}}}

\newcommand{\G}{{\mbox{\tiny G}}}

\newcommand{\BH}{{\mbox{\tiny BH}}}

\newcommand{\BL}{{\mbox{\tiny BL}}}

\newcommand{\K}{{\mbox{\tiny K}}}

\newcommand{\ext}{\mathrm{ext}}
\newcommand{\inter}{\mathrm{int}}

\newcommand{\mrm}{\mathrm}
\begin{document}
%\title{Toward a Better Understanding of How Compact Stars Approach Black Holes} 
%\title{Toward a Better Understanding of Universality as Compact Stars Approach Black Holes} 
\title{I-Love-Q Relations: From Compact Stars to Black Holes} 

\author{Kent Yagi}
%\email{kyagi@physics.montana.edu}
\address{Department of Physics, Princeton University, Princeton, New Jersey 08544, USA}
\address{eXtreme Gravity Institute, Department of Physics, Montana State University, Bozeman, Montana 59717, USA}

\author{Nicol\'as Yunes}
\address{eXtreme Gravity Institute, Department of Physics, Montana State University, Bozeman, Montana 59717, USA}

%\author{Anton B. Vorontsov}
%\affiliation{Department of Physics, Montana State University, Bozeman, MT 59717, USA.}

\date{\today}

%%%%%%%%%%%%%%%%%%%%%%%%%%%%%%%%%%%%%%%%%%%%%%%%%
\begin{abstract} 

%EoS-dependent relations for NSs and QSs
The relations between most observables associated with a compact star, such as the mass and radius of a neutron star or a quark star, typically depend strongly on their unknown internal structure. 
%
%Universal I-Love-Q relations
The recently discovered \emph{I-Love-Q} relations (between the moment of inertia, the tidal deformability and the quadrupole moment) are however approximately insensitive to this structure.
%
%BHs have their own relations
These relations become exact for stationary black holes in General Relativity as shown by the no-hair theorems, mainly because black holes are vacuum solutions with event horizons.  
%
%General strategy/goals of our research program.
In this paper, we take the first steps toward studying how the approximate I-Love-Q relations become exact in the limit as compact stars become black holes. 
%
%Anisotropic compact stars as a toy model to see whether the compact star relations reach the BH relations. Why anisotropy
To do so, we consider a toy model for compact stars, i.e.~incompressible stars with anisotropic pressure, which allows us to model an equilibrium sequence of stars with ever increasing compactness that approaches the black hole limit arbitrarily closely. 
%
%How we proceed
We numerically construct such a sequence in the slow-rotation and in the small-tide approximations by extending the Hartle-Thorne formalism, and then extract the I-Love-Q trio from the asymptotic behavior of the metric tensor at spatial infinity.
%
%Numerical results
We find that the I-Love-Q relations approach the black hole limit in a nontrivial way, with the quadrupole moment and the tidal deformability changing sign as the compactness and the amount of anisotropy are increased.
%Divergence and Maclaurin-like spheroids
Through a generalization of Maclaurin spheroids to anisotropic stars, we show that the multipole moments also change sign in the Newtonian limit as the amount of anisotropy is increased because the star becomes prolate.
%
%Analytic results
We also prove analytically that the stellar moment of inertia reaches the black hole limit as the compactness reaches a critical black hole value in the strongly anisotropic limit. 
%
%non-GR theories
Modeling the black hole limit through a sequence of anisotropic stars, however, can fail when considering other theories of gravity. 
%dCS
We calculate the scalar dipole charge and the moment of inertia in a particular parity-violating modified theory and find that these quantities do not tend to their black hole counterparts as the anisotropic stellar sequence approaches the black hole limit. 

\end{abstract}


%04.30.Db Wave generation and sources
% 04.50.Kd Modified theories of gravity
% 04.25.-g Approximation methods; equations of motion
%04.25.Nx Post-Newtonian approximation; perturbation theory; related approximations
%97.60.Jd Neutron stars

\maketitle

%\tableofcontents

%%%%%%%%%%%%%%%%%%%%%%%%%%%%%%%%%%%
\section{Introduction}

%NSs, QSs, hybrid stars, internal structure, nuclear physics
A plethora of compact stars with masses between 1 $M_{\odot}$ and 2 $M_\odot$ and with radii of approximately $12$ km have been discovered through a variety of astrophysical observations~\cite{steiner-lattimer-brown,Lattimer:2013hma,Miller:2013tca,Ozel:2015fia}. The limited accuracy of these observations, coupled to degeneracies in the observables with respect to different models for the nuclear physics at supranuclear densities encoded in the equation of state (EoS), have prevented observations from elucidating the internal structure of compact objects. For example, X-ray observations do not typically allow us to confidently state whether the compact objects observed are standard neutron stars~\cite{lattimer_prakash2001,lattimer-prakash-review,ozel-review,Lattimer:2012nd}, or hybrid stars with quark-gluon plasma cores~\cite{Alford:2004pf,Alford:2015dpa}, or perhaps even strange quark stars~\cite{SQM}. Future observations of compact objects could shed some light on this problem, as the accuracy of the observations increases and more observables are obtained~\cite{2012SPIE.8443E..13G,Psaltis:2013fha}. 

%I-Love-Q, 
The extraction of information from these future observations is aided by the use of approximately universal relations, i.e.~relations between certain observables that are roughly insensitive to the EoS~\cite{I-Love-Q-Science,I-Love-Q-PRD,Baubock:2013gna,Psaltis:2013zja,Psaltis:2013fha}.  For example, the moment of inertia $I$, the tidal deformability $\lambda_2$ (or tidal \emph{Love} number) and the (rotation-induced) quadrupole moment $Q$ satisfy relations (the so-called \emph{I-Love-Q} relations) that are EoS insensitive to a few~\% level~\cite{I-Love-Q-Science,I-Love-Q-PRD}. Such relations are useful to analytically break degeneracies in the models used to extract information from X-ray and gravitational-wave observations of compact objects. This information, in turn, allows us to better probe nuclear physics~\cite{Psaltis:2013fha} and gravitational physics~\cite{I-Love-Q-Science,I-Love-Q-PRD}. 

%No-hair
Similar universal relations exist among the multipole moments of compact stars~\cite{Pappas:2013naa,Stein:2014wpa,Yagi:2014bxa,Chatziioannou:2014tha,Majumder:2015kfa}, i.e.~the coefficients of a multipolar expansion of the gravitational field far from the compact object. These \emph{no-hair like} relations resemble the well-known, black hole (BH) no-hair relations of general relativity (GR)~\cite{robinson,israel,israel2,hawking-uniqueness0,hawking-uniqueness,carter-uniqueness,Gurlebeck:2015xpa}. The latter state that all multipole moments of an uncharged, stationary BH in GR can be prescribed only in terms of the first two (the BH mass and spin). The no-hair like relations of compact stars differ from the BH ones in that the former require knowledge of the first three stellar multipole moments to prescribe all higher moments in a manner that is roughly insensitive to the underlying EoS. 

%Main questions to address, some difficulties
But how are the approximate I-Love-Q and no-hair like relations for compact stars related to those that hold for BHs exactly? One way to address this question is to carry out simulations of compact stars that gravitationally collapse into BHs, extract the I-Love-Q and multipole moments and study how the relations evolve dynamically. However, not only are such simulations computationally expensive, but the machinery employed in the past would no longer be useful, as it is valid only for stationary spacetimes, i.e.~the \emph{Geroch-Hansen} multipole moments~\cite{geroch,hansen} used for example in~\cite{I-Love-Q-Science,I-Love-Q-PRD,Pappas:2013naa,Yagi:2014bxa} are not well-defined for non-stationary spacetimes. One would have to employ a dynamical generalization of these moments and develop a procedure to extract them from dynamical simulations.   

%Anisotropic stars
A simpler way to gain some insight is to consider how the universal relations evolve in a sequence of equilibrium stellar configurations\footnote{Another approach is to consider ``BH mimickers'' whose compactness can reach that of BHs, such as the gravastars considered in~\cite{Pani:2015tga}. The latter, however, are very different from neutron stars or quark stars.} of ever increasing compactness that approaches the compactness of BHs arbitrarily closely. Such a sequence, however, cannot be constructed from neutron star solutions with isotropic pressure, as used in the original I-Love-Q~\cite{I-Love-Q-Science,I-Love-Q-PRD} and no-hair like relations~\cite{Pappas:2013naa,Yagi:2014bxa}; such stars have a maximum stellar compactness (i.e.~the ratio between the stellar mass and radius) that is well below the BH limit. An alternative approach is to consider a sequence of anisotropic stars\footnote{Anisotropic stars are here only used as a toy model to study an equilibrium sequence of compact stars that can reach the BH limit, and not as a realistic model for compact stars.} (see e.g.~\cite{1997PhR...286...53H} for a review of anisotropic stars), which, for example, in the Bowers and Liang (BL) model~\cite{1974ApJ...188..657B} can reach BH compactnesses for incompressible stars in the strongly anisotropic limit. 

%Relating follicly-challenged compact stars to bald BHs
Following this logic, in~\cite{Yagi:2015upa} we studied how the no-hair like relations for compact stars approach the BH limit. We first showed that the stellar shape transitions from prolate to oblate as the compactness is increased. We then showed that the multipole moments approach the BH limit with a power-law scaling and that the no-hair like relations also approach the BH limit in a very nontrivial way.  In this paper we extend these investigations in a variety of ways and clarify several points that were left out of the initial analysis. 

First, in this paper we consider both slowly-rotating stars and tidally-deformed stars, which allow us to study how the I-Love-Q relations approach the BH relations in the BH limit. In~\cite{Yagi:2015hda}, we constructed tidally-deformed or slowly-rotating, anisotropic compact stars to third order in spin for various realistic EoSs. We here follow~\cite{Yagi:2015hda} but focus on incompressible stars, as this allows us to construct an equilibrium sequence of anisotropic stars that approaches the BH limit arbitrarily closely. 

%What's new compared to the letter (II): Analytic calculations
Second, we extend the analysis of~\cite{Yagi:2015upa} by carrying out analytic calculations in various limits: (i) the weak-field limit, (ii) beyond the weak-field limit, (iii) the strong-field limit and (iv) the strongly-anisotropic limit. In the first limit, we expand all equations in small compactness and retain only the leading terms in the expansion. This leads to anisotropic stars modeled as incompressible spheroids with arbitrary rotation that reduce to Maclaurin spheroids~\cite{1969efe..book.....C,1983bhwd.book.....S,2014grav.book.....P} in the isotropic limit.  When going beyond the weak-field limit, we retain subleading terms in the small compactness expansion, which is equivalent to a post-Minkowskian (PM) expansion; we extend the work of~\cite{Chan:2014tva} for isotropic stars to the anisotropic case and derive the moment of inertia and tidal deformability. In the third limit, we expand all equations about the maximum compactness allowed for incompressible stars, extending the analysis of~\cite{damour-nagar} to anisotropic stars and deriving the tidal deformability for specific choices of the anisotropy parameter. In the fourth limit, we expand all equations about the maximum anisotropy allowed by the BL model, analytically deriving the moment of inertia for incompressible stars as a function of the compactness.

%What's new compared to the letter (III): non-GR theories
Third, we study whether an equilibrium sequence of anisotropic compact stars can be used to study the BH limit of stellar observables in theories other than GR. As an example, we work in dynamical Chern-Simons (dCS) gravity~\cite{jackiw,Smith:2007jm,CSreview}, a parity violating modified theory of gravity that is motivated from string theory~\cite{polchinski2}, loop quantum gravity~\cite{alexandergates,taveras,calcagni} and effective theories of inflation~\cite{Weinberg:2008hq}. We treat this modified theory as an effective field theory and assume that the GR deformation is small. Such a treatment ensures the well-posedness of the initial value problem~\cite{Delsate:2014hba}. Slowly-rotating, anisotropic compact stars to linear order in spin in dCS gravity were constructed in~\cite{Yagi:2015hda} using realistic EoSs within the anisotropy model proposed by Horvat \textit{et al}.~\cite{Horvat:2010xf}. We now extend the treatment in~\cite{Yagi:2015hda} to the BL anisotropy model and focus on the incompressible case.

%-------------------------------------------
\subsection{Executive Summary}

\begin{figure}[htb]
\begin{center}
\includegraphics[width=8.5cm,clip=true]{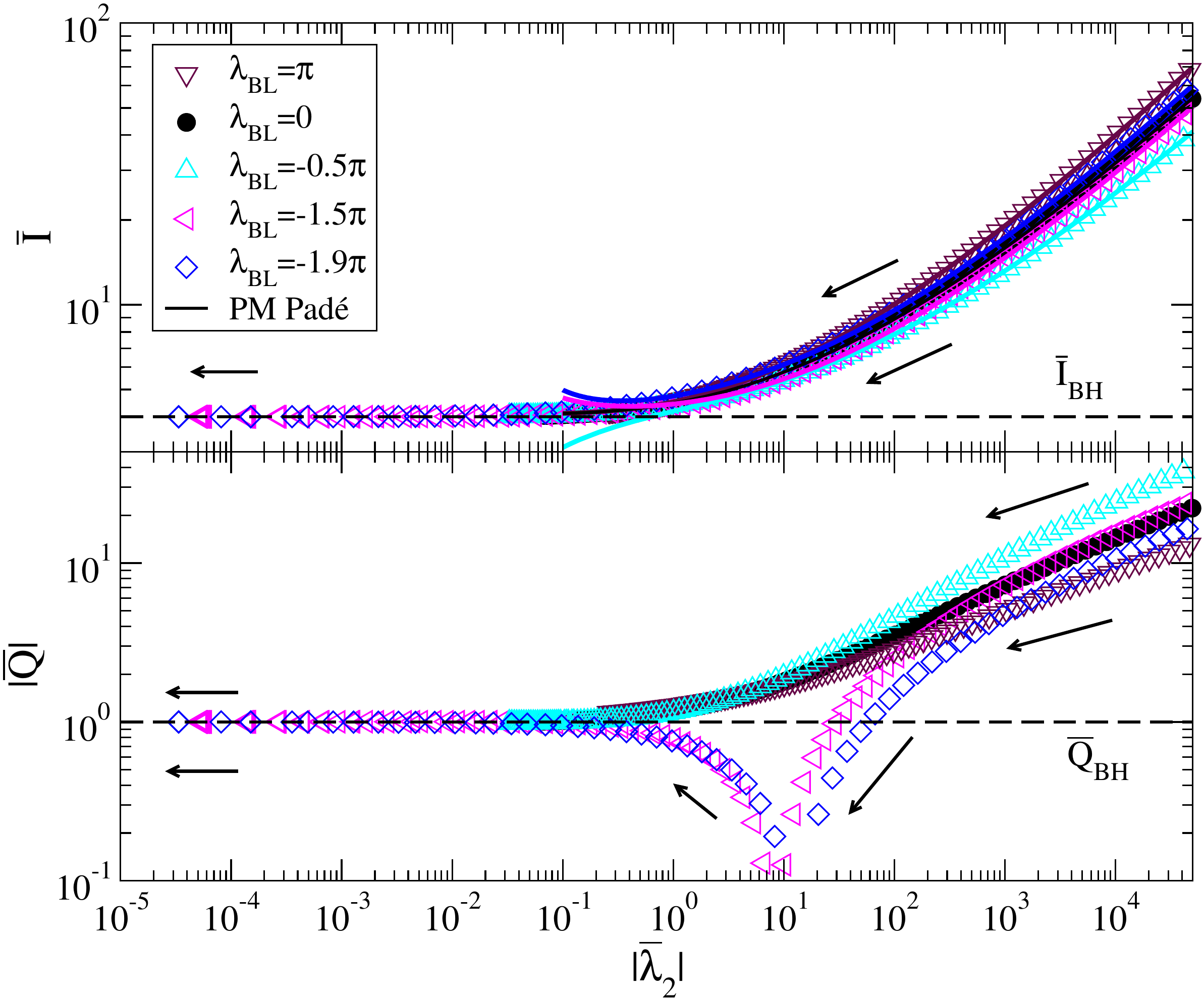}  
\caption{\label{fig:univ-polyn0} (Color online) Relations between the dimensionless moment of inertia $\bar I \equiv I/M_*^3$ and the dimensionless tidal deformability $\bar \lambda_2 \equiv \lambda_2/M_*^5$ (top), and between the dimensionless quadrupole moment $\bar Q \equiv - Q/(M_*^3 \chi^2)$ and $\bar \lambda_2$ (bottom) for an equilibrium sequence of anisotropic, incompressible stars with varying compactness (the arrows indicate increasing compactness), given some anisotropy parameter $\lambda_\BL$, to leading-order in slow rotation and tidal deformation. $M_*$ is the stellar mass, $\chi \equiv J/M_*^2$ is the dimensionless spin parameter, with $J$ the magnitude of the stellar spin angular momentum. Isotropic stars correspond to $\lambda_\BL=0$, while strongly-anisotropic stars correspond to $\lambda_\BL = - 2\pi$. Our numerical results are validated by analytic PM calculations (solid curves). The dashed horizontal lines correspond to the BH values of $\bar I$ and $\bar Q$, while $\bar \lambda_{2,\BH} = 0$ is the BH value for the dimensionless tidal deformability. Observe that the I-Love-Q relations of anisotropic stars approach the BH limit continuously.  
}
\end{center}
\end{figure}

%I-Love-Q in GR (Fig.1)
Let us now present a brief summary of our results. We find that the I-Love-Q relations for strongly anisotropic stars in GR indeed approach the BH limit as one increases the compactness. Figure~\ref{fig:univ-polyn0} shows evidence for this by presenting the I-Love-Q relations for incompressible stars with a variety of anisotropy parameters $\lambda_{\BL}$ in the BL model~\cite{1974ApJ...188..657B}. The isotropic case is recovered when $\lambda_\BL=0$, while $\lambda_\BL = - 2 \pi$ corresponds to the strongly anisotropic limit. The BH limit ($\bar \lambda_{2,\BH}=0$) corresponds to $\bar I_\BH = 4$ and $\bar Q_\BH = 1$, shown with dashed horizontal lines. We confirm the validity of our numerical results by comparing them to an analytic calculation of the I-Love relations in the PM approximation (solid curves in the top panel of Fig.~\ref{fig:univ-polyn0}). Observe that the relations approach the BH limit as the compactness is increased (shown with arrows) in a way that depends quite strongly on $\lambda_\BL$, with $\bar \lambda_2$ and $\bar Q$ changing sign as the BH limit is approached. 

We also find that the approach of the I-Love-Q relations to the BH limit appears to be \emph{continuous}, as shown in Fig.~\ref{fig:univ-polyn0}. That is, we find no evidence of the discontinuity hypothesized in~\cite{Glampedakis:2013jya}, based on a weak-field calculation of the quadrupole moment of strongly anisotropic, incompressible stars. We in fact prove analytically that the moment of inertia of a strongly anisotropic, incompressible compact star reaches the BH limit continuously as the compactness is increased. We do so by constructing slowly-rotating, anisotropic incompressible stars to linear order in spin in the strongly anisotropic limit ($\lambda_\BL = -2\pi$) and analytically deriving $\bar I$ as a function of the compactness $C$ in terms of hypergeometric functions. Taylor expanding $\bar I$ about $C_\BH = 1/2 + {\cal{O}}(\chi^{2})$, with $\chi$ the dimensionless spin parameter, we find that $\bar I(C) = \bar I_\BH + \mathcal{O}(C-C_\BH,\chi^{2})$. 

%Maclaurin-like spheroids
The quadrupole moment changes sign as it approaches the BH limit, as shown in Fig.~\ref{fig:univ-polyn0}, but is this the case for all multipole moments?  We find that this is not the case by constructing incompressible spheroids with anisotropic pressure and arbitrary rotation in the weak-field limit. We derive a necessary condition on the anisotropy model such that spheroidal configurations are realized and find that the BL model satisfies such a condition. We then calculate the $\ell$th mass and current multipole moments, $M_\ell$ and $S_\ell$, in the slow-rotation limit within the BL model and find that $M_{2\ell + 2}$ and $S_{2\ell + 3}$ are both proportional to $1/(4\pi + 5 \lambda_\BL)^{\ell+1}$. This means that the sign of only ($M_2$, $M_6$, $M_{10}$...) and ($S_3$, $S_7$, $S_{11}$...) is opposite to that of the isotropic case when $\lambda_\BL < -4\pi/5$, which is consistent with the results of~\cite{Glampedakis:2013jya} for the quadrupole moment $M_2$. In particular, the sign of ($M_4$, $M_8$, $M_{12}$...) and ($S_5$, $S_9$, $S_{13}$...) is the same as that of the sign of the isotropic case even in the strongly-anisotropic limit. 

\begin{figure}[htb]
\begin{center}
\includegraphics[width=8.5cm,clip=true]{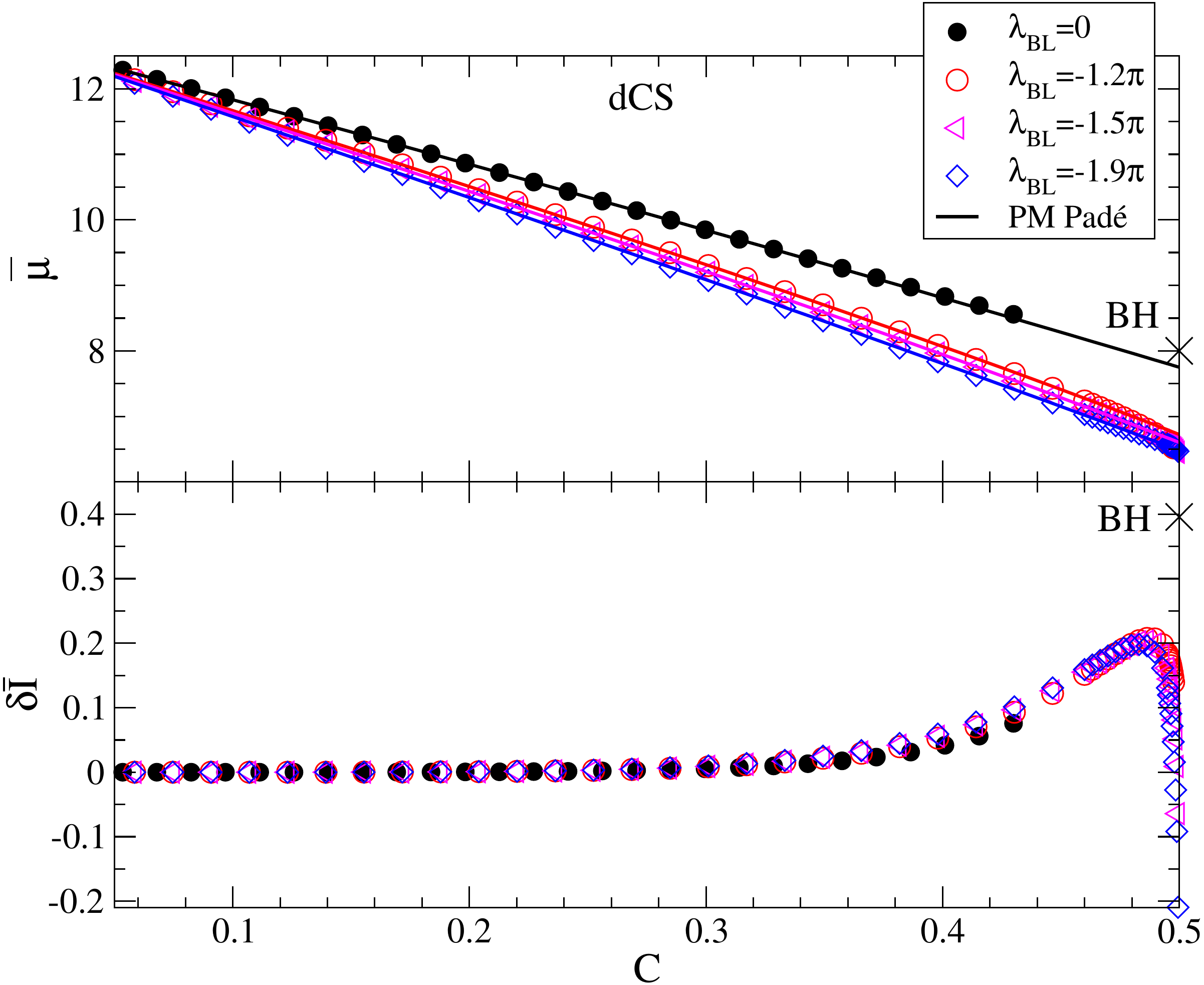}  
\caption{\label{fig:I-Love-Shen-error} (Color online) Dimensionless scalar dipole charge $\bar \mu$ [Eq.~\eqref{eq:dipole}] (top) and the dCS correction to the dimensionless moment of inertia $\delta \bar I$ (normalized by the dimensionless dCS coupling constant and the GR value of $\bar I$) [Eq.~\eqref{eq:delta-Ibar}] (bottom) for a sequence of anisotropic, incompressible stars labeled by stellar compactness $C$ and anisotropy parameter $\lambda_\BL$. Corresponding BH values are shown by black crosses. Our numerical results are validated by analytic PM calculations (solid curves) [Eq.~\eqref{eq:22Pade}] in the top panel. Observe that, unlike in the GR case, $\bar \mu$ and $\delta \bar I$ do not approach the BH limit as one decreases $\lambda_\BL$ and increases $C$.}
\end{center}
\end{figure}

%scalar charge and I in dCS (Fig.2)
Although the I-Love-Q relations for compact stars approach the BH limit as one increases the compactness in GR, we find that this is not always the case in other theories of gravity when the limit is modeled through an equilibrium sequence of anisotropic stars. Figure~\ref{fig:I-Love-Shen-error} presents evidence for this by showing the scalar dipole charge and the correction to the dimensionless moment of inertia in dCS gravity as a function of the stellar compactness. Once more, we validate our numerical results by comparing them to analytic PM relations for the scalar dipole charge. Observe that unlike in the GR case, these quantities do not approach the dCS BH limit (shown with black crosses) as one increases the compactness. This result suggests that modeling the BH limit through strongly anisotropic stars is not appropriate in certain modified theories of gravity. 

%Organization
The remainder of this paper presents the details of the calculations that led to the results summarized above. In Sec.~\ref{sec:formalism}, we explain the formalism that we use to construct slowly-rotating and tidally-deformed anisotropic stars. We also describe the BL anisotropy model and show how the maximum stellar compactness for a non-rotating configuration approaches the BH one in the strongly anisotropic limit. In Sec.~\ref{sec:analytic}, we present analytic calculations of the stellar moment of inertia, tidal deformability and multipole moments in certain limits. %In Sec.~\ref{sec:transition}, we explain how the stellar shape changes from prolate to oblate as one increases the compactness. We also provide a simple explanation for why rotating, strongly anisotropic stars in the weak-field limit become prolate with a Maclaurin-like spheroid model. 
In Sec.~\ref{sec:BH-limit}, we present numerical results that show how the I-Love-Q relations approach the BH limit in GR. We also show that the scalar dipole charge and the correction to the moment of inertia in dCS gravity do not approach the BH limit. Finally, in Sec.~\ref{sec:future}, we give a short summary and discuss various avenues for future work. We use the geometric units of $c=1=G$ throughout this paper.

%%%%%%%%%%%%%%%%%%%%%%%%%%%%%%%%%%%
\section{Formalism and Anisotropy Model}
\label{sec:formalism}

In this section, we first explain the formalism we use to construct slowly-rotating or tidally-deformed compact stars with anisotropic pressure and extract the stellar multipole moments and tidal deformability. We then explain the specific anisotropic model that we will use throughout the paper. We present the spherically-symmetric background solution and describe the maximum compactness such a solution can possess for polytropic EoSs of the form $p = K \rho^{1 + 1/n}$. Here $p$ and $\rho$ are the stellar radial pressure and energy density, while $K$ and $n$ are constants. Henceforth, the stellar compactness is defined by $C \equiv M_*/R_*$, where $M_*$ and $R_*$ are the stellar mass and radius for a non-rotating configuration respectively. 

%---------------------
\subsection{Formalism}

Let us first explain how one can construct slowly-rotating compact stars with anisotropic pressure, by following~\cite{1976ApJ...207..279Q,1999ApJ...520..788K,benhar,Yagi:2015hda} and extending the Hartle-Thorne approach~\cite{hartle1967,Hartle:1968ht} to third order in spin. Let us assume the spacetime is stationary and axisymmetric, such that the metric can be written as
\begin{align}
\label{Eq:metric-slow-rot}
ds^2 &= - e^{\nu (r)} \left[ 1 + 2 \epsilon^2 h(r,\theta) \right] dt^2 +  e^{\lambda (r)} \left[ 1 + \frac{2 \epsilon^2 m(r,\theta)}{r - 2 M(r)} \right] dr^2 
\nn \\
&+  r^2 \left[ 1 + 2 \epsilon^2 k(r,\theta)  \right] 
\left( d\theta^2  + \sin^2 \theta \left\{ d\phi 
- \epsilon \left[ \Omega - \omega (r,\theta)  + \epsilon^2 w (r,\theta) \right] dt \right\}^2 \right) + \mathcal{O}(\epsilon^4)\,,
\end{align}
where $\nu$ and $\lambda$ are functions of the radial coordinate $r$ only, while $\omega$, $h$, $k$, $m$ and $w$ are functions of both $r$ and $\theta$. The quantity $\epsilon$ is a book-keeping parameter that labels the order of an expression in $(M_{*} \Omega)$, where $\Omega$ is the spin angular velocity. The surface is defined as the location where the radial pressure vanishes. We transform the radial coordinate via
\be
\label{eq:rtoR}
r(R, \theta) = R + \epsilon^2 \xi(R,\theta) + \mathcal{O}(\epsilon^4)\,,
\ee
so that the spin perturbation to the radial pressure and density vanish throughout the star~\cite{hartle1967,Hartle:1968ht}. The enclosed mass function $M(r)$ is defined via
\be
e^{- \lambda(r)}  \equiv 1 - \frac{2 M(r)}{r}\,,
\ee
and thus, $M_{*}$ is the value of $M(r)$ evaluated at the stellar surface $R_{*}$. We decompose $\omega$, $h$, $k$, $m$, $\xi$ and $w$ in Legendre polynomials~\cite{Yagi:2015hda}.

The stress-energy tensor for matter with anisotropic pressure can be written as~\cite{Doneva:2012rd,Silva:2014fca,Yagi:2015hda}
\be
\label{eq:matter}
T_{\mu \nu} = \rho \; u_\mu u_\nu + p \; k_\mu k_\nu + q \; \Pi_{\mu \nu}\,,
\ee
where $q$ is the tangential pressure and $u^\mu$ is the fluid four-velocity, given by $u^\mu = (u^0, 0,0,\epsilon \, \Omega \, u^0)$, with $u^0$ determined through the normalization condition $u^\mu u_\mu = -1$.  $k^\mu$ is a unit radial vector that is spacelike ($k^\mu k_\mu = 1$) and orthogonal to the four-velocity ($k^\mu u_\mu=0$) of the fluid, while $\Pi_{\mu \nu} \equiv g_{\mu \nu} + u_\mu u_\nu - k_\mu k_\nu$ is a projection operator onto a two-surface orthogonal to $u^\mu$ and $k^\mu$. We introduce the anisotropy parameter $\sigma \equiv p-q$~\cite{Doneva:2012rd,Silva:2014fca} with $\sigma = 0$ corresponding to isotropic matter. Following the treatment of metric perturbations, we expand $\sigma$ in the slow-rotation approximation and decompose each term in Legendre polynomials as
\ba
\label{eq:sigma-Leg}
\sigma(R,\Theta) = \sigma_0^{(0)} (R) + \epsilon^2 \left\{ \sigma^{(2)}_{0}(R) + \sigma^{(2)}_{2}(R) P_2(\cos\theta) \right\}  +  \mathcal{O}(\epsilon^4)\,.
\ea
Notice that the superscript (subscript) in $\sigma_\ell^{(n)}$ corresponds to the order of the spin (Legendre) decomposition. The function $\sigma_0^{(0)}$ needs to be specified \emph{a priori}, and it in fact defines the anisotropy model. The function $\sigma^{(2)}_{2}$ is determined consistently by solving the perturbed Einstein equations, once $\sigma_0^{(0)}$ is chosen~\cite{Yagi:2015hda}. The function $\sigma^{(2)}_{0}$ is irrelevant in this paper as it only affects the stellar mass at subleading order in a small spin expansion. %quadratic order in spin and we here calculate all multipole moments to leading order in spin.
 
We construct slowly-rotating compact star solutions with anisotropic pressure as follows. First, we substitute the metric ansatz and the matter stress-energy tensor mentioned above into the Einstein equations. We then expand in small spin (or equivalently in $\epsilon$) and solve the perturbed Einstein equations order by order in $\epsilon$. In the interior region, we solve the equations numerically with a regularity condition at the center. In the exterior region, we solve the equations analytically with an asymptotic flatness condition at spatial infinity. We finally match the two solutions at the stellar surface to determine any integration constants. The latter determine the moment of inertia $I$, the quadrupole moment $Q$ and the octupole moment $S_3$ of the exterior solution at linear, quadratic and third order in spin respectively.

In this paper, we also construct non-rotating but tidally-deformed compact stars to extract the stellar tidal deformability~\cite{damour-nagar,binnington-poisson}. We are here particularly interested in the quadrupolar, electric-type tidal deformability, $\lambda_2$, which is defined as the ratio of the tidally-induced quadrupole moment and the external tidal field strength. We follow~\cite{hinderer-love,damour-nagar,binnington-poisson} and treat tidal deformations as small perturbations of an isolated compact star solution. Such a tidally-deformed compact star can be constructed similarly to how we construct slowly-rotating solution, except that we set $\Omega = \omega = w =0$, as we are only interested in electric-type, even-parity perturbations.

For convenience, we work with the following dimensionless quantities throughout:
\be
\bar I \equiv \frac{I}{M_*^3}\,, \quad \bar Q \equiv - \frac{Q}{M_*^3 \chi^2}\,, \quad \bar S_3 \equiv - \frac{S_3}{M_*^4 \chi^3}\,, \quad \bar \lambda_2 \equiv \frac{\lambda_2}{M_*^5}\,.
\ee
Here, the dimensionless spin parameter $\chi$ is defined through the magnitude of the spin angular momentum $J$ by $\chi \equiv J/M_*^2$, with $J$ only kept to $\mathcal{O}(\epsilon)$. The BH value of each dimensionless quantity above is $\bar I_\BH = 4$~\cite{membrane}, $\bar Q_\BH = 1$~\cite{hansen}, $\bar S_{3,\BH}=1$~\cite{hansen} and $\bar \lambda_{2,\BH}=0$~\cite{damour-nagar,binnington-poisson,kol-smolkin,chakrabarti1,Gurlebeck:2015xpa}. We here choose to work with the above choice of normalization introduced in~\cite{I-Love-Q-Science,I-Love-Q-PRD,Pappas:2013naa,Stein:2014wpa,Yagi:2014bxa}, but clearly this choice is not unique. In fact, one can choose other normalizations, for example involving the stellar compactness, which may improve the universality in the I-Love-Q relations and no-hair like relations for compact stars among stellar multipole moments~\cite{Majumder:2015kfa}. Other choices of normalization, nonetheless, will not affect the conclusions we arrive at in this paper.

%---------------------
\subsection{Anisotropy Model}
\label{sec:ani-model}

Let us now describe the specific anisotropy model that we use in this paper. Following BL~\cite{1974ApJ...188..657B}, we choose
\be
\label{eq:BL}
\sigma_0^{(0)} (R) =  \frac{\lambda_\BL}{3} (\rho + 3p) (\rho + p) \left( 1-\frac{2M}{R} \right)^{-1} R^2\,.
\ee
Here, $\lambda_\BL$ is a constant parameter that characterizes the amount of anisotropy. Isotropic pressure corresponds to $\lambda_\BL = 0$, since then both $\sigma^{(0)}_{0}$ and $\sigma^{(2)}_{2}$ vanish. The particular form of $\sigma_{0}^{(0)}$ in Eq.~\eqref{eq:BL} was proposed such that the Tolman-Oppenheimer-Volkoff equation could be solved analytically for a spherically-symmetric, incompressible (polytropic index $n=0$, i.e.~$\rho = {\rm{const.}}$) anisotropic star to yield~\cite{1974ApJ...188..657B}
\allowdisplaybreaks
\begin{align}
\label{eq:M-n0}
M(R) &= \frac{C}{R_*^2}R^3\,, \\
\label{eq:rho-n0}
\rho (R) &= \frac{3}{4\pi} \frac{C}{R_*^2}\,, \\
\label{eq:p-n0}
p(R) &=  -\frac{3}{4\pi} \frac{C}{R_*^2} \frac{(1-2 C)^{\gamma} - (1-2 C R^2/R_*^2)^{\gamma}}{3 (1-2 C)^{\gamma} - (1-2 C R^2/R_*^2)^{\gamma}}\,, \nn \\
\\
\label{eq:nu-n0}
\nu (R) &= \frac{1}{\gamma} \ln \left[ \frac{3 (1-2 C)^\gamma - (1-2 C R^2/R_*^2)^\gamma}{2} \right]\,,
\end{align}
where $\gamma$ is defined by
\be
\label{eq:gamma}
\gamma \equiv \frac{1}{2} \left( 1 + \frac{\lambda_\BL}{2\pi} \right) \,.
\ee

\begin{figure}[htb]
\begin{center}
\includegraphics[width=8.cm,clip=true]{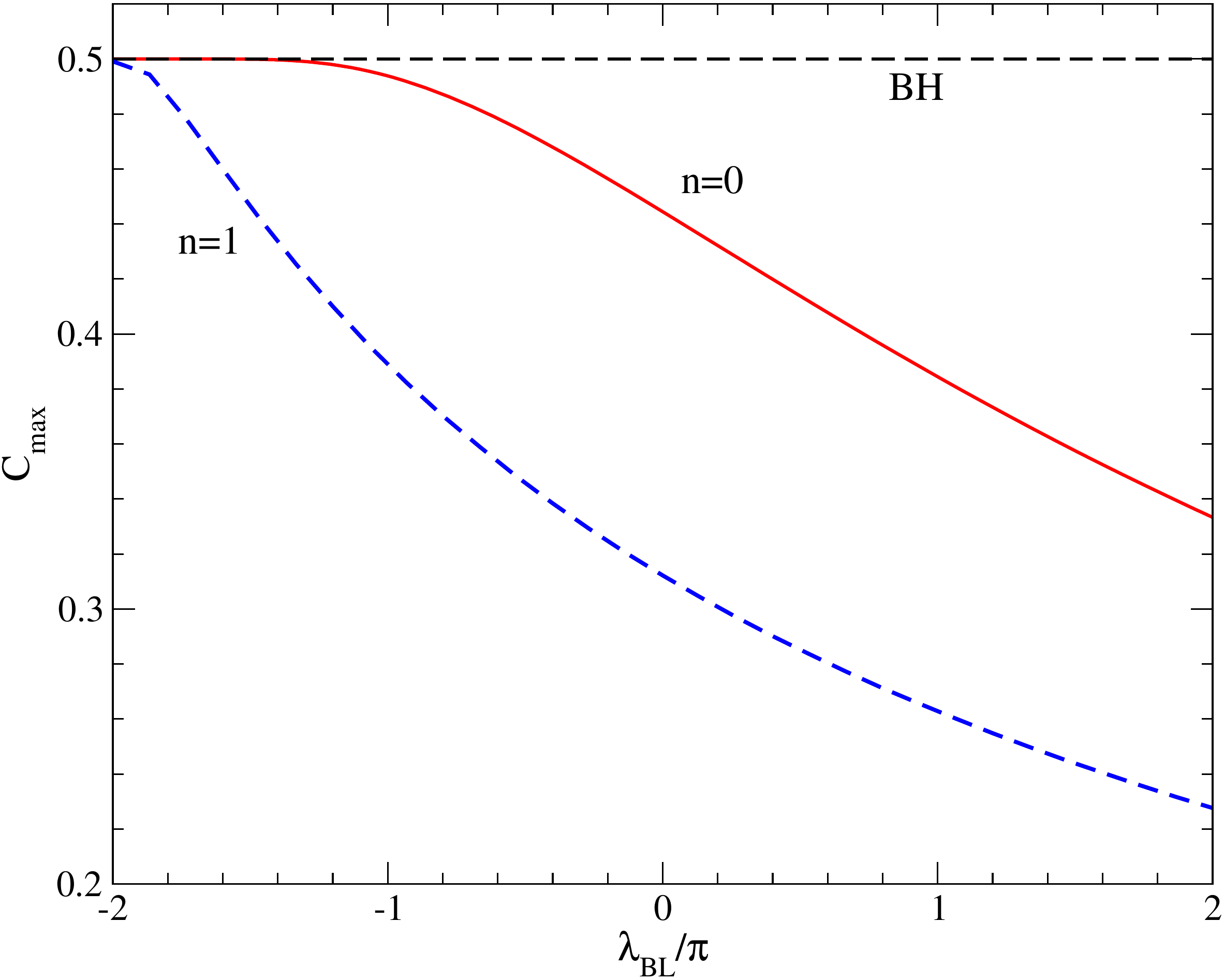}  
\caption{\label{fig:C-max-BL-n0-noH} (Color online) Maximum compactness realized for the $n=0$ (solid) and $n=1$ (dashed) polytopes as a function of the anisotropy parameter $\lambda_\BL$. The horizontal dashed line corresponds to the compactness of a non-rotating BH.
}
\end{center}
\end{figure}

How do the maximum compactness $C_{\max}$ of anisotropic stars differ from the isotropic case? 
One can find $C_{\max}$ for anisotropic stars by finding the value of $C$ for which the central radial pressure $p(R=0)$ diverges~\cite{1974ApJ...188..657B}:
\be
\label{eq:max-comp}
C_{\max} = \frac{1}{2} \left( 1 - 3^{-1/\gamma} \right)\,.
\ee
Clearly, the solution in Eqs.~\eqref{eq:p-n0} and~\eqref{eq:nu-n0} is only well-defined for $\lambda_{\BL} \geq - 2 \pi$, such that $C_{\max} \geq 0$. Such a condition on $\lambda_\BL$ also ensures that $p\geq 0$ and the solution does not diverge when $C \leq C_{\max}$. The red solid curve in Fig.~\ref{fig:C-max-BL-n0-noH} presents $C_{\max}$ as a function of $\lambda_\BL$ for incompressible stars [Eq.~\eqref{eq:max-comp}]. Observe that in the isotropic incompressible case, $C_{\max} = 4/9 \approx 0.444...$, while $C_{\max}$ approaches $1/2$, the compactness of a non-rotating BH, in the $\lambda_\BL \to -2\pi$ limit. This is exactly why we consider anisotropic stars in this paper, as they allow us to construct a sequence of equilibrium stars that approaches the BH limit arbitrarily closely. Henceforth, $C_{\BH} \equiv 1/2$, the compactness of a non-rotating BH, since we work to leading order in the slow rotation approximation and $C_{\Kerr} = 1/2 + {\cal{O}}(\chi^{2})$.  For reference, Fig.~\ref{fig:C-max-BL-n0-noH} also shows the maximum compactness for anisotropic stars with an $n=1$ polytropic EoS (blue dashed curve), which is always smaller than that of incompressible stars.

\begin{figure*}[htb]
\begin{center}
\includegraphics[width=5.cm,clip=true]{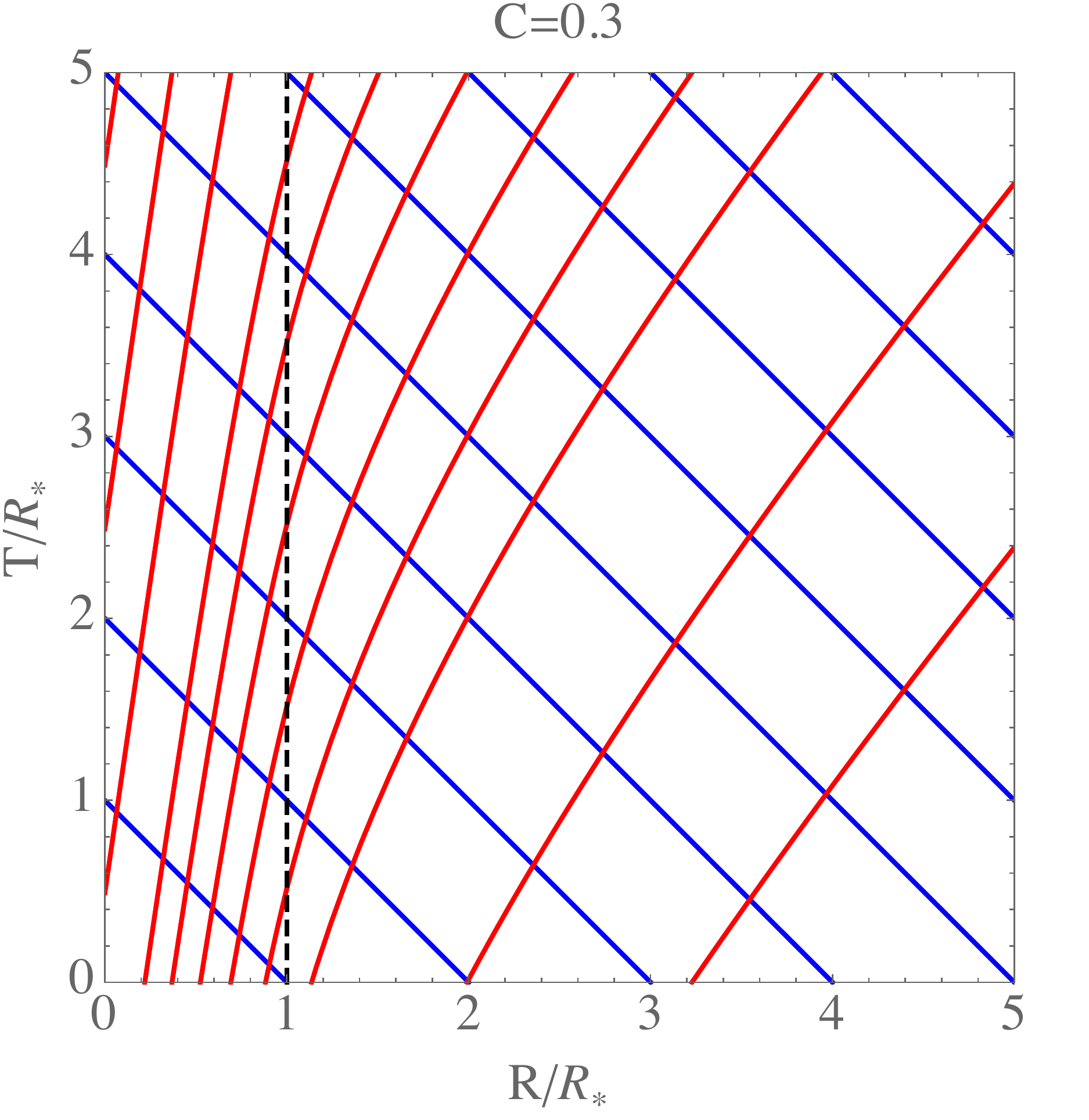}  
\includegraphics[width=5.cm,clip=true]{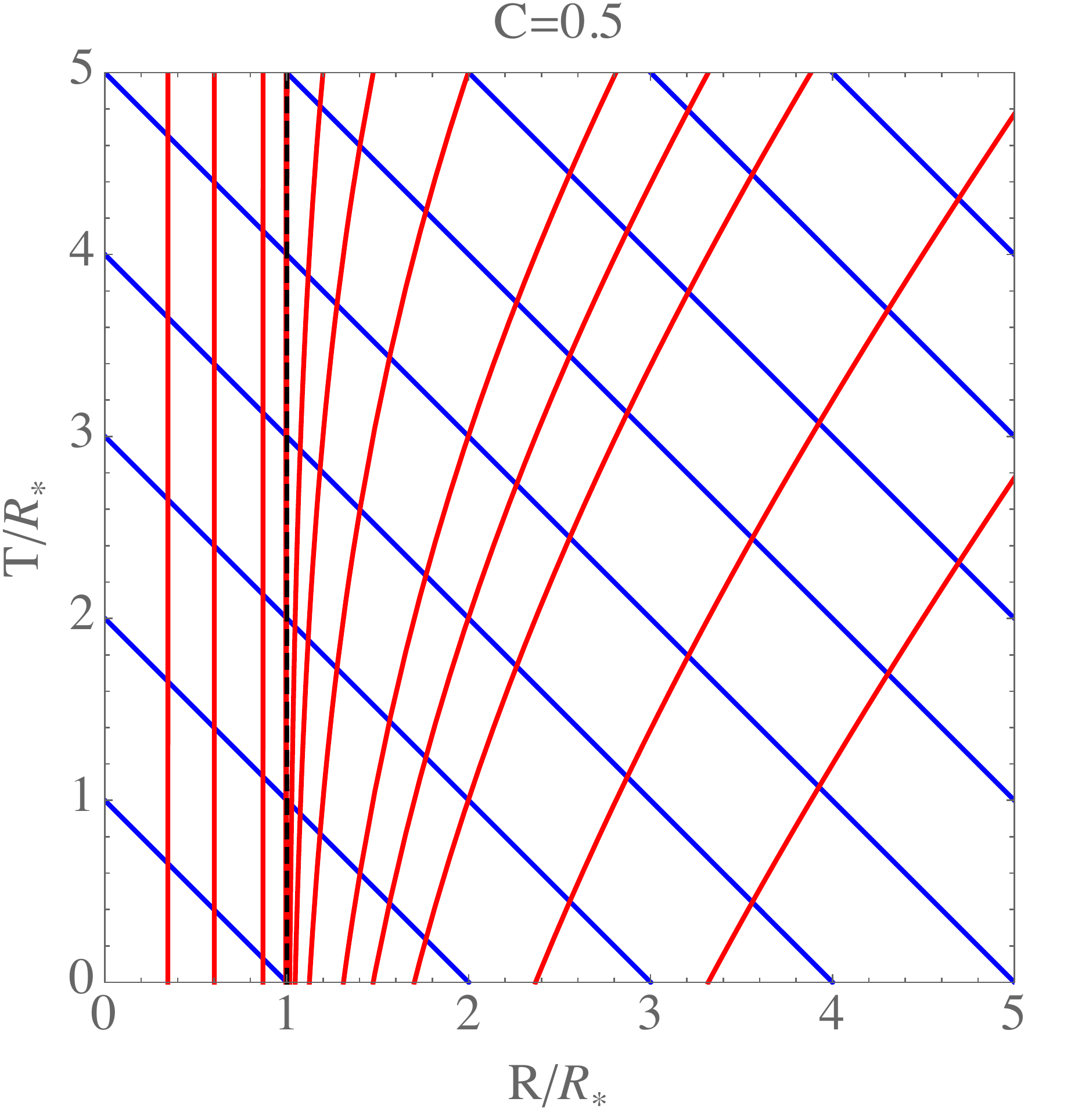}  
\includegraphics[width=5.cm,clip=true]{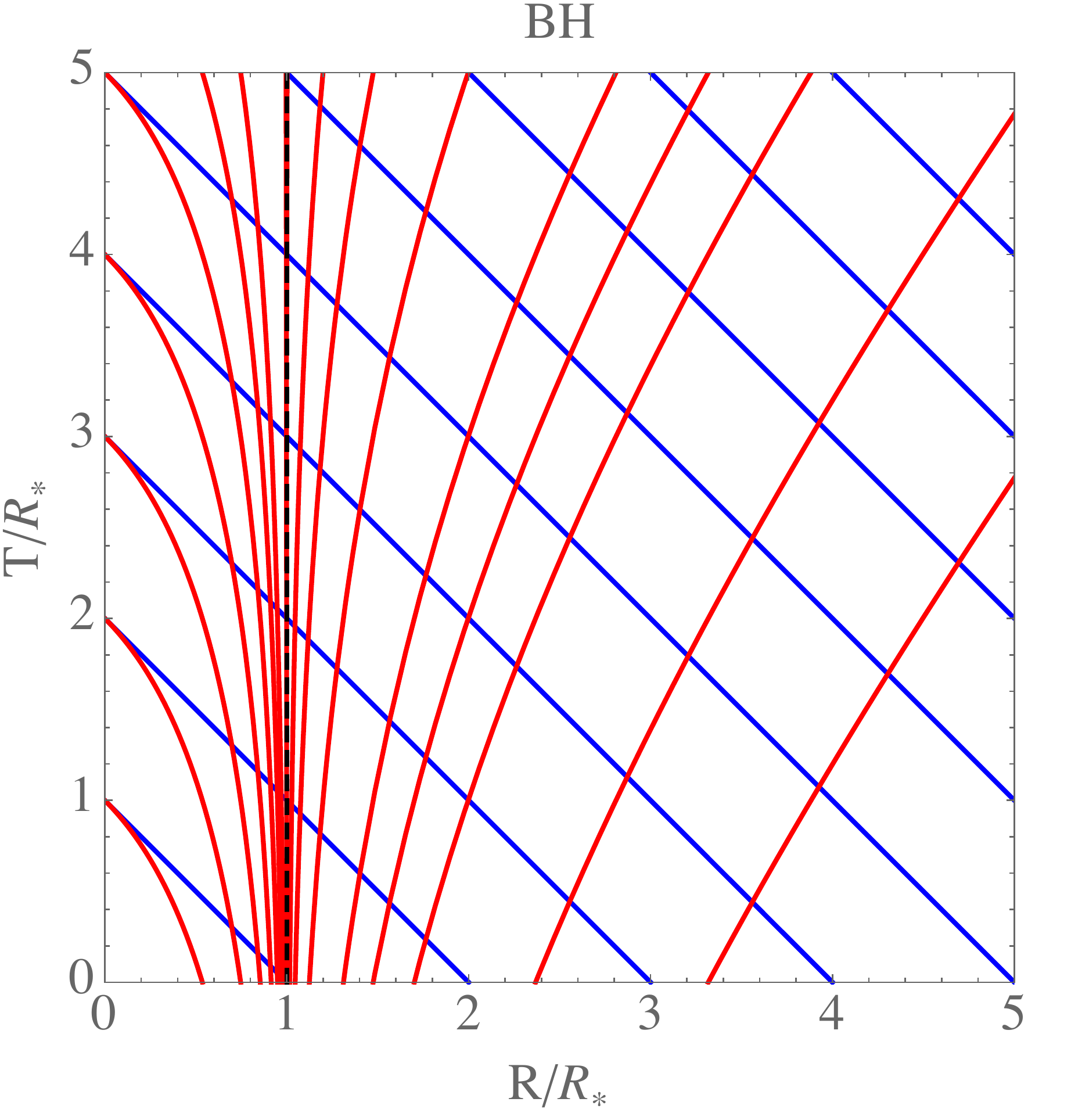}  
\caption{\label{fig:causal} (Color online) Causal structure of a non-rotating compact anisotropic star with $C=0.3$ (left) and $C=0.5$ (middle) with $\lambda_\BL=-2\pi$ and a non-rotating BH (right). Blue and red curves correspond to ingoing and outgoing null geodesics respectively. The opening angle between these curves at each crossing point shows that of a light cone at each point. Observe that the surface of an anisotropic star with $C=0.5$ is a trapped surface and radiation inside the star cannot escape to outside. Observe also how the causal structure of the interior region for such a star is different from that of a BH.
}
\end{center}
\end{figure*}

Although the maximum compactness of non-rotating, anisotropic stars can reach the compactness of non-rotating BHs in the strongly anisotropic limit ($\lambda_\BL \to -2\pi$), the causal structure inside such a star is quite different from that of a BH. The left, middle and right panels of Fig.~\ref{fig:causal} show the causal structure of non-rotating anisotropic compact stars with $C=0.3$ and $C=0.5$, and that of a BH respectively. To construct these panels, we introduce a new (retarded) time coordinate $T = v-R$~\cite{Quinzacara:2012zz}, where $v = t + r_*$ is a null coordinate with $r_{*}$ the tortoise coordinate in the exterior and interior regions, given by Eqs.~\eqref{eq:tortoise-ext} and~\eqref{eq:tortoise-int} respectively. The ingoing null geodesics (blue lines in the figure) are given by $v= \mrm{const.}$, while the outgoing null geodesics (red curves in the figure) are given by $t - r_* = \mrm{const.}$ The opening angle between blue and red curves at each point represents that of the light cone, while the stellar surface or the event horizon are denoted by a black dashed vertical line. Observe that a photon emitted inside a star with $C=0.3$ can escape out to spatial infinity, while that from a star with $C=0.5$ cannot. This is because the surface for the latter acts as a trapped surface, just like the event horizon of a BH. However, notice that the causal structure in the interior region between an anisotropic compact star with $C=0.5$ and a BH is different. In particular, a photon emitted inside an anisotropic star with $C=0.5$ stays at a constant radius $R$, while the one  emitted inside a BH eventually falls into singularity.

%%%%%%%%%%%%%%%%%%%%%%%%%%%%%%%%%%%
\section{Analytic Calculations}
\label{sec:analytic}

Before diving into a full numerical analysis, let us first present some analytic calculations of the multipole moments, of $\bar I$ and of $\bar \lambda_2$ for incompressible, anisotropic stars in certain limits. Later on, in Sec.~\ref{sec:BH-limit}, we will use these analytic calculations to verify the validity of our numerical analysis. In Sec.~\ref{sec:weak}, we calculate multipole moments with arbitrary $\ell$ in the weak-field (or so-called ``Newtonian'') limit by constructing a spheroid with arbitrary rotation that reduces to Maclaurin spheroids~\cite{1969efe..book.....C,1983bhwd.book.....S,2014grav.book.....P} in the isotropic limit. In Sec.~\ref{sec:PM}, we calculate $\bar I$ and $\bar \lambda_2$ using a PM analysis, which is valid beyond the weak-field limit. In Sec.~\ref{sec:tidal-strong}, we derive $\bar \lambda_2$ in the strong-field, or maximum-compactness limit, for specific choices of $\lambda_\BL$, while in Sec.~\ref{sec:strong-ani} we calculate $\bar I$ as a function of $C$ in the strongly anisotropic limit, $\bar \lambda\BL = - 2 \pi$. In the latter, we prove that $\bar{I}$ approaches the moment of inertia of a non-spinning BH in the limit as the compactness goes to $1/2$. 

%----------------------------
\subsection{Weak-field Limit}
\label{sec:weak}

We here derive the multipole moments of anisotropic, incompressible stars in the weak-field limit. We begin by constructing an anisotropic stellar solution that is spheroidal and valid to arbitrary order in rotation. In the isotropic case, such a solution reduces to Maclaurin spheroids~\cite{1969efe..book.....C,1983bhwd.book.....S,2014grav.book.....P}. We then use the anisotropic spheroidal solution to find the multipole moments of the star. We conclude this subsection by providing a phenomenological explanation for why rotating strongly-anisotropic stars are prolate in the Newtonian limit, and whether such anisotropic stars are stable to perturbations in the amount of anisotropy.

%-------------------
\subsubsection{Maclaurin-like Spheroids}

Let us first prove that $\sigma$ is purely a function of $r$ in axisymmetry, working in spherical coordinates $(r,\theta,\phi)$. The $r$ and $\theta$ components of the hydrostatic equilibrium equation are given by~\cite{Glampedakis:2013jya}
\begin{align}
\label{eq:Euler-r}
\frac{1}{\rho} \frac{\partial p}{\partial r} + \frac{\partial \Phi}{\partial r} + \frac{1}{\rho} \frac{2 \sigma}{r} &= 0\,, \\
\label{eq:Euler-theta}
\frac{1}{\rho} \frac{\partial p}{\partial \theta} + \frac{d \Phi}{\partial \theta} - \frac{1}{\rho} \frac{\partial \sigma}{\partial \theta} &= 0\,, 
\end{align}
where $\Phi = \Phi_\G + \Phi_c$ with $\Phi_\G$ and $\Phi_c$ representing the gravitational and centrifugal potentials respectively. We decompose $\bm B = (p, \Phi, \sigma)$ using Legendre polynomials $P_\ell$ as
\be
\bm B = \sum_\ell \bm B_\ell (r) P_\ell (\cos \theta)\,.
\ee
Substituting this into Eqs.~\eqref{eq:Euler-r} and~\eqref{eq:Euler-theta}, one finds  
\begin{align}
\label{eq:Euler-r-2}
\frac{1}{\rho} \frac{d p_\ell}{d r} + \frac{d \Phi_{\ell}}{d r} + \frac{1}{\rho} \frac{2 \sigma_\ell}{r} &= 0  \quad (\ell \geq 0)\,, \\
\label{eq:Euler-theta-2}
\frac{1}{\rho}  p_\ell + \Phi_{\ell} - \frac{1}{\rho} \sigma_{\ell} &= 0  \quad (\ell > 0)\,, 
\end{align}
while Eq.~\eqref{eq:Euler-theta} is automatically satisfied when $\ell = 0$. Taking a derivative of Eq.~\eqref{eq:Euler-theta-2} with respect to $r$ and combining this with Eq.~\eqref{eq:Euler-r-2}, one finds
\be
\frac{d \sigma_\ell}{d r} + \frac{2 \sigma_\ell}{r} = 0 \quad (\ell > 0)\,.
\ee
Imposing regularity at the stellar center, one finds $\sigma_\ell = 0$ for $\ell > 0$. This is consistent with~\cite{Glampedakis:2013jya}, in which the authors showed that $
\sigma_2 = 0$ at quadratic order in spin. Since the only non-vanishing Legendre mode of $\sigma$ is the $\ell=0$ mode, $\sigma$ cannot depend on any angular coordinates for stationary and axisymmetric, incompressible and anisotropic stars in the weak-field limit. 

Let us now derive a necessary condition for $\sigma_0 (r)$ such that the star is a spheroid. This condition comes from the equations of structure that determine the  gravitational potential, $\Phi_\G$, and the radial pressure $p$. The Poisson equation determines the former, which is not modified from its form in the isotropic case:
\be
\label{eq:Poisson}
\left( \frac{\partial^2}{\partial x^2} +  \frac{\partial^2}{\partial y^2} +  \frac{\partial^2}{\partial z^2} \right) \Phi_\G = 4 \pi \rho\,,
\ee
working in Cartesian coordinates $(x,y,z)$, with the $z$-axis identified with the axis of rotation. The hydrostatic equilibrium equation determines $p(x,y,z)$, and with $\sigma = \sigma_0 (r)$ its components in the $x$ and $z$ directions are
\begin{align}
\label{eq:hydro-x}
\frac{1}{\rho} \frac{\partial p}{\partial x} &= - \frac{\partial \Phi_\G}{\partial x} - \frac{1}{\rho} \frac{2 \sigma_0}{r} n^x + \Omega^2 x\,,  \\
\label{eq:hydro-z}
\frac{1}{\rho} \frac{\partial p}{\partial z} &= - \frac{\partial \Phi_\G}{\partial z} - \frac{1}{\rho} \frac{2 \sigma_0}{r} n^z\,, 
\end{align}
where $r = |\bm{x}| = \sqrt{x^2 + y^2 + z^2}$ and $n^i = x^i/r$. The second term on the right-hand side in Eqs.~\eqref{eq:hydro-x} and~\eqref{eq:hydro-z} correspond to an extra force induced by pressure anisotropy.

We now solve the above set of differential equations to find a condition on $\sigma_{0}(r)$. Equation~\eqref{eq:Poisson} can be easily solved using Green's function as in the isotropic case, and the solution for a spheroid is given by~\cite{1983bhwd.book.....S,2014grav.book.....P}
\be
\label{eq:Phi-answer}
\Phi_\G (x,y,z) = -\pi \rho \left[ A_0 a_1^2 -A_1 (x^2 + y^2) - A_3 z^2 \right]\,.
\ee
Here, $A_0$, $A_1$ and $A_3$ are given in terms of the stellar eccentricity $e^2 (\equiv 1 - a_3^2 / a_1^2)$ by 
\begin{align}
A_0 &= {\frac {2 \sqrt {1 - {e}^{2}}\arcsin \left( e \right) }{{e}}}\,, \\
A_1 &= {\frac {\sqrt {1 - {e}^{2}}\arcsin \left( e \right) }{{e}^{3}}}-{\frac{1 - {e}^{2}}{{e}^{2}}}\,, \\
A_3 &= \frac{2}{e^2} - 2\,{\frac {\sqrt {1 - {e}^{2}}\arcsin \left( e \right) }{{e}^{3}}}\,,
\end{align}
where the stellar radius on the $x$ (or $y$) and $z$ axes are denoted by $a_1$ and $a_3$ respectively. Note that $0<e^2<1$ when a star is oblate, while $e^2 < 0$ when a star is prolate, in which case $a_{1}$ ($a_{3}$) do not correspond to the semi-major (semi-minor) radius of the star. Imposing that the solution is spheroidal implies that the radial pressure $p$ must be given by~\cite{1983bhwd.book.....S}
\be
\label{eq:p-anz}
p (x,y,z) = p_c \left( 1 - \frac{x^2+y^2}{a_1^2} - \frac{z^2}{a_3^2} \right)\,,
\ee
where $p_c$ is the central radial pressure. Substituting Eqs.~\eqref{eq:Phi-answer} and~\eqref{eq:p-anz} into Eqs.~\eqref{eq:hydro-z} and~\eqref{eq:hydro-x}, one finds that $\sigma_0$ must satisfy
\be
\label{eq:sigma-spheroid}
\sigma_0 = \left( \frac{p_c}{a_3^2} - \pi A_3 \rho^2 \right) r^2\,, 
\ee
while $\Omega^{2}$ must satisfy
\be
\label{eq:Omega2-spheroid}
\Omega^2 = \frac{\left[ 2 \pi \rho^2 (A_1 - A_3) a_3^2 + 2 p_c \right] a_1^2 - 2 a_3^2 p_c}{\rho a_1^2 a_3^2}\,.
\ee
Equation~\eqref{eq:sigma-spheroid} shows that $\sigma$ needs to be proportional to $r^2$ to realize a spheroid, since recall that for an incompressible fluid $\rho = {\rm{const.}}$

%--------------------------------------------
\subsubsection{Spheroidal Shape and Multipole Moments}

Consider now a specific choice of $\sigma_{0}$ that is consistent with the BL model and that satisfies Eq.~\eqref{eq:sigma-spheroid}: 
\be
\label{eq:sigma-0-choice}
\sigma_0 = \frac{\lambda_\BL}{3} \rho^2 r^2\,.
\ee
Note that if one transforms Eq.~\eqref{eq:sigma-0-choice} to the $R$ coordinate with Eq.~\eqref{eq:rtoR}, Eq.~\eqref{eq:sigma-0-choice} agrees with Eq.~\eqref{eq:BL} in the Newtonian limit at zeroth order in spin. The radial part of the $\ell=2$ contribution at second-order in spin agrees with $\sigma_2^{(2)} (R)$ in Eq.~\eqref{eq:sigma-Leg} in the Newtonian limit. Combining Eq.~\eqref{eq:sigma-0-choice} with Eq.~\eqref{eq:sigma-spheroid}, one finds 
\be
\label{eq:pc-BL}
p_c = \left( \pi A_3 + \frac{\lambda_\BL}{3} \right) \rho^2 a_3^2\,.
\ee
Substituting this into Eq.~\eqref{eq:Omega2-spheroid}, one finds the relation between $\Omega$ and $e^2$ as
\be
\label{eq:Omega}
\Omega^2 = \frac{3}{2 \pi} \frac{\Omega_\K^2}{\sqrt{1-e^2}} \left\{ \pi \left[ A_1- \left( 1-e^2 \right) A_3 \right] + \frac{ \lambda_\BL}{3} e^2\right\}\,,
\ee
where 
\be
\label{eq:OmegaK}
\Omega_\K^2 \equiv \frac{M_*}{a_1^3} = \frac{4 \pi}{3} \rho \sqrt{1-e^2}
\ee
corresponds to the (squared) Keplerian angular velocity at the equatorial surface. 

\begin{figure*}[htb]
\begin{center}
\includegraphics[width=7.5cm,clip=true]{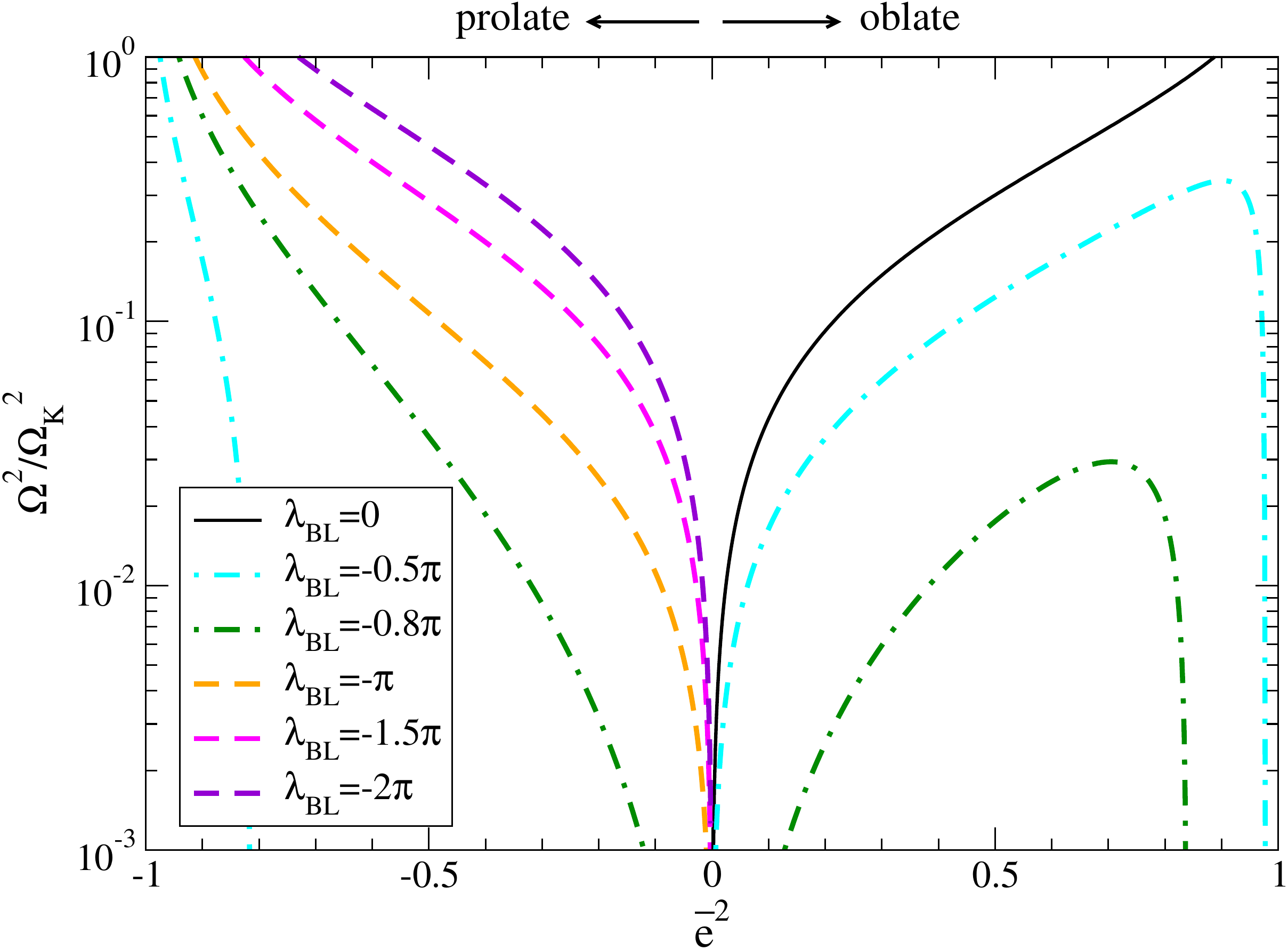}  
\quad
\includegraphics[width=7.cm,clip=true]{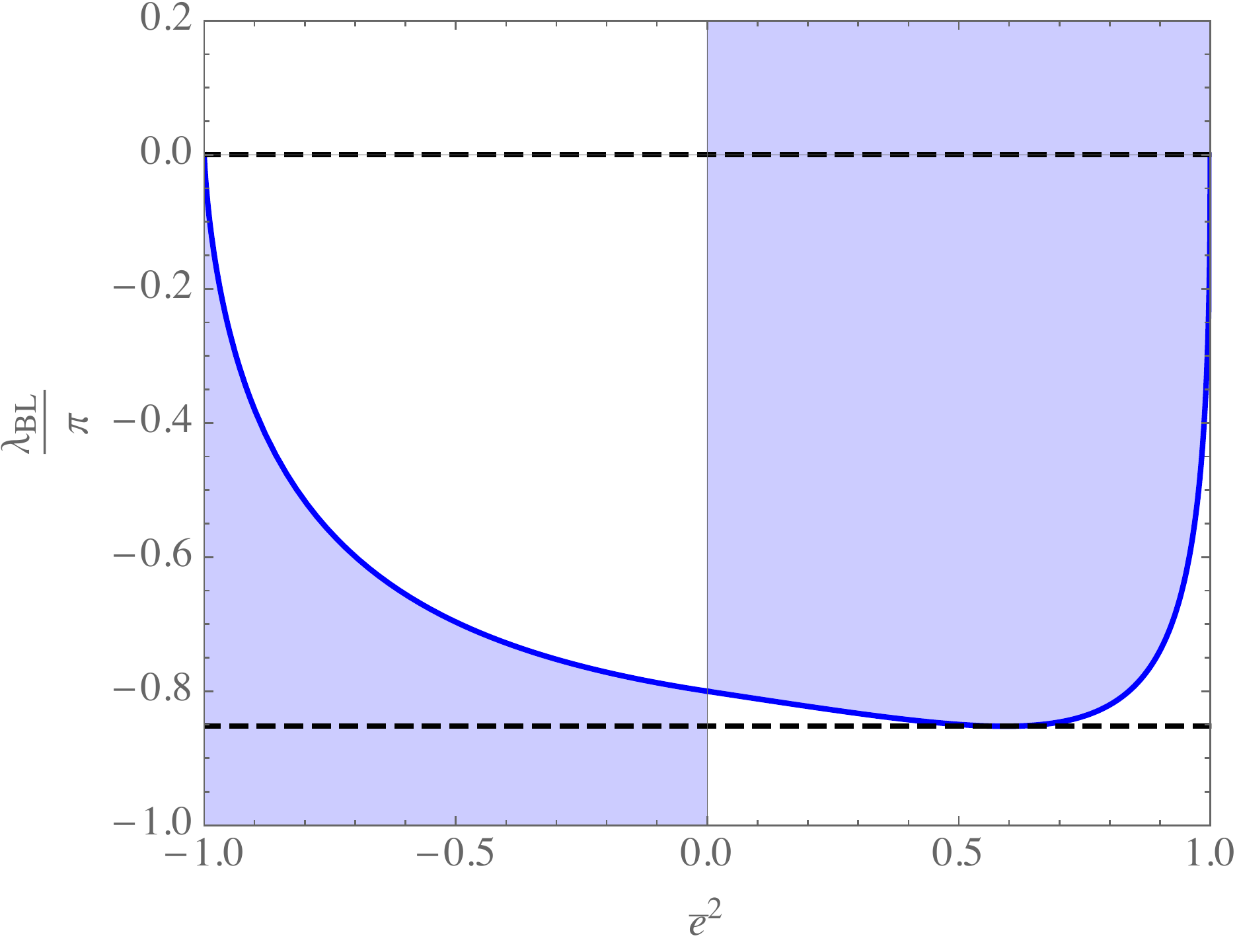}  
\caption{\label{fig:Omega2-e2} (Color online) (Left) $\Omega^2/\Omega_\K^2$ [Eq.~\eqref{eq:Omega}] as a function of $\bar e^2$ [Eq.~\eqref{eq:ebar2}] for various $\lambda_\BL$. The solid and dashed curves correspond to the oblate and prolate configurations respectively. The dotted-dashed curve can have both oblate and prolate configurations. (Right) Regions in the $\lambda_\BL$--$\bar e^2$ plane where $\Omega^2 \geq 0$ and the stellar solution exists (shaded). Observe that when $\lambda_\BL>0$ ($\lambda_\BL\lesssim 0.85 \pi$) only oblate (prolate) configurations exist. Both prolate and oblate branches exist when $\lambda_\BL \in (-0.85\pi,0)$.
}
\end{center}
\end{figure*}

The left panel of Fig.~\ref{fig:Omega2-e2} shows $\Omega^2/\Omega_\K^2$ as a function of $\bar e^2$ in Eq.~\eqref{eq:Omega} for various values of $\lambda_\BL$. Here, $\bar e^2$ is defined as
\be
\label{eq:ebar2}
\bar e^2 = 
\begin{cases}
   1 - \frac{a_3^2}{a_1^2} = e^2 & (a_1 \geq a_3)\,, \\
    \frac{a_1^2}{a_3^2}-1 = \frac{e^2}{1-e^2}  & (a_1 < a_3)\,,
\end{cases}
\ee
such that the star is prolate (oblate) for $-1 < \bar e^2 < 0$ ($0 < \bar e^2 < 1$).
The right panel of Fig.~\ref{fig:Omega2-e2} shows regions in the $\lambda_\BL$--$\bar e^2$ plane in which $\Omega^2$ is positive (i.e.~the stellar solution exists). Observe that when $\lambda_{\BL} \geq 0$ the stellar shape can only be oblate, while when $\lambda_{\BL} \lesssim -0.85\pi$ it can only be prolate. Observe also that the shape can be either oblate or prolate when $-0.85\pi \lesssim \lambda_\BL < 0$, but one of the branches is probably unstable.

We are now in a position to derive the multipole moments for such an anisotropic spheroidal solution, following the analysis in~\cite{Stein:2014wpa}. The mass and current multipole moments for a Newtonian incompressible star with arbitrary rotation are given by 
\begin{align}
\label{eq:M2lp2}
M_{2\ell+2} &= {\frac {\left( -1 \right) ^{\ell+1} 3}{ \left( 2\,\ell+3 \right)  \left( 2\,\ell+5 \right) }} \, \Omega_\K^2 \,{e}^{2\,\ell+2} \, {a_1}^{2\,\ell+
5}\,, \\
\label{eq:S2lp1}
S_{2\ell+1} &={\frac {\left( -1 \right) ^{\ell} 6}{ \left( 2\,\ell+3 \right)  \left( 2\,\ell+5 \right) }} \,\Omega_\K^2 \, \Omega\, {e}^{2\,\ell}\,{a_1}^{2
\,\ell+5}\,. 
\end{align}
One could numerically invert Eq.~\eqref{eq:Omega} and insert the solution into the equations above in order to express the multipole moments entirely in terms of the angular velocity.

Let us instead consider the slow-rotation approximation, as this allows us to invert Eq.~\eqref{eq:Omega} analytically. One can perturbatively solve Eq.~\eqref{eq:Omega} for $e^2$ within the slow-rotation approximation to find
\be
\label{eq:e2}
e^2 = {\frac {10 \pi}{ 4\,\pi +5\,\lambda_\BL }} \left( \frac{\Omega}{\Omega_\K} \right)^2 + \mathcal{O} \left[ \left( \frac {\Omega}{\Omega_\K} \right)^{4} \right]\,.
\ee
Notice that the star is oblate ($e^{2}>0$) when $\lambda_{\BL} > -4 \pi/5$, while it is prolate ($e^{2} < 0$) when $\lambda_{\BL} < -4\pi/5$. Notice, however, that there is a range of $\lambda_{\BL}$ in which one cannot construct an equilibrium configuration within the small-rotation approximation, because then $e^{2} > 1$, which would force one of the semi-axes to be imaginary. This occurs in the range $-4 \pi/5 < \lambda_{\BL} <  \lambda_\BL^{(e^2=1)}$, where we have defined
\be
\label{eq:lambda-e1}
\lambda_\BL^{(e^2=1)} = - \frac{4\pi}{5} \left( 1 - \frac{5}{2} \frac{\Omega^2}{\Omega_\K^2} \right)\,.
\ee
Notice also that when $e^{2} \gg -1$ ($|\lambda_{\BL} + 4\pi/5| \ll 1$ and $\lambda_{\BL} < - 4\pi/5 $), the star remains prolate and the semi-axes can remain real.

Let us now evaluate the multipole moments in the slow-rotation approximation. Substituting Eq.~\eqref{eq:e2} into Eqs.~\eqref{eq:M2lp2} and~\eqref{eq:S2lp1} and eliminating $\rho$ using $\Omega_\K^2$ in Eq.~\eqref{eq:OmegaK}, one finds
\begin{align}
\allowdisplaybreaks
\label{eq:M2lp2-2}
M_{2\ell+2} &= \frac { \left( -1 \right) ^{\ell+1} \, 3 \,{(10 \pi)}^{\ell+1}} {\left( 2\,\ell+3 \right)  \left( 2\,\ell+5 \right) \left( 4\,\pi +5\,\lambda_\BL
 \right)^{\ell+1}  } \left( \frac {\Omega}{\Omega_\K} \right) ^{2 \ell+2} \Omega_K^2 \, a_1^{2\ell+5} + \mathcal{O} \left[ \left( \frac {\Omega}{\Omega_\K} \right)^{2 \ell+4} \right]\,, \nn \\ 
 \\
\label{eq:S2lp1-2}
S_{2\ell+1} &= \frac { \left( -1 \right) ^{\ell} \, 6 \,{(10 \pi)}^{\ell}}{\left( 2\,\ell+3 \right)  \left( 2\,\ell+5 \right)  \left( 
4\,\pi +5\,\lambda_\BL \right)^\ell}  \left( \frac {\Omega}{\Omega_\K} \right) ^{2 \ell + 1}  \Omega_\K^3 \, a_1^{2\ell+5} + \mathcal{O} \left[ \left( \frac {\Omega}{\Omega_\K} \right)^{2 \ell+3} \right]\,. 
\end{align}
Since $M_{2\ell + 2} \propto (4\pi + 5 \lambda_\BL)^{-(\ell+1)}$, the sign of $M_{2\ell + 2}$ is opposite to that in the isotropic case for $\ell$ even, i.e.~the moments $M_2$, $M_6$, $M_{10}$..., or simply $M_{4 \ell +2}$ with integer $\ell$, flip sign when $\lambda_\BL < -4\pi/5$. Similarly, since $S_{2\ell + 1}\propto (4\pi + 5 \lambda_\BL)^{-\ell}$, the sign of $S_{2\ell + 1}$ is opposite to that in the isotropic case when $\ell$ is odd, i.e.~the moments $S_3$, $S_7$, $S_{11}$..., or simply $S_{4 \ell +3}$ with integer $\ell$, flip sign when $\lambda_{\BL} < -4\pi/5$. In the quadrupole case, $M_2 \propto (4\pi + 5 \lambda_\BL)^{-1}$, which is consistent with~\cite{Glampedakis:2013jya}. 

Before proceeding, notice that the multipole moments in Eqs.~\eqref{eq:M2lp2-2} and~\eqref{eq:S2lp1-2} diverge at $\lambda_\BL = -4\pi/5$. This divergence originates from Eq.~\eqref{eq:e2} and is an artifact of the slow-rotation approximation. Such an approximation breaks down near the divergence, and such a feature is absent in the multipole moments valid for arbitrary rotation in Eqs.~\eqref{eq:M2lp2} and~\eqref{eq:S2lp1}. For example, the left panel of Fig.~\ref{fig:Omega2-e2} shows that $e^2$ (or $\bar e^2$) is finite for a given $\Omega^2/\Omega_\K^2$ at $\lambda_\BL = -4\pi/5$, and hence, the multipole moments in Eqs.~\eqref{eq:M2lp2} and~\eqref{eq:S2lp1} are also finite at $\lambda_\BL = -4\pi/5$. We will see how different manifestations of this divergence contaminate our calculations later on; fortunately, it does not affect the behavior of the multipole moments near the BH limit, since the latter requires we take the $\lambda_{\BL} \to -2\pi$ limit, which is far from the divergent value of $\lambda_{\BL}$.

%--------------------
\subsubsection{From Oblate to Prolate Anisotropic Stars}
\label{sec:shape_analy}

Let us now try to develop a better understanding of why the stellar shape changes from oblate to prolate as one decreases $\lambda_\BL$. In particular, the goal of this subsection is to understand the behavior of Eq.~\eqref{eq:e2} from a force balance argument on a fluid element inside an isolated star. 

We begin by defining the sum of the pressure gradient and potential gradient of a fluid element (normalized by $\Omega_\K^2$ for later convenience) acting along the $x$ axis as
\be
F_x \equiv - \frac{1}{\Omega_\K^2} \left( \frac{1}{\rho} \frac{\partial p}{\partial x} + \frac{\partial \Phi_\G}{\partial x} \right)\,,
\ee
and decompose it within the small eccentricity approximation ($|e^2| \ll 1$) as
\be
\label{eq:F-small-ecc}
F_x = \sum_{k=0} F_x^{(k)} \delta^{2k}\,,
\ee
where $\delta$ is a book keeping parameter that counts orders in $e$. 

Let us now look at force balance along the $x$-axis at second-order in rotation. At this order, the second term on the right hand side of Eq.~\eqref{eq:hydro-x} vanishes (as it is spin independent), and hence the force balance equation is given by
\be
\label{eq:force-balance-x}
F_{x,0}^{(1)} + F_{x,\lambda_\BL}^{(1)} + \frac{\Omega^2}{\Omega_\K^2} x = 0\,,
\ee
where $F_{x,0}^{(1)}$ corresponds to $F_x^{(1)}$ for an isotropic star while $F_{x,\lambda_\BL}^{(1)}$  represents its anisotropic correction for a fixed $e^2$. 
From the exact solution of $p$ and $\Phi_\G$ given in Eqs.~\eqref{eq:Phi-answer} and~\eqref{eq:p-anz}, one finds 
\be
\label{eq:fx0}
F_{x,0}^{(1)} = - \frac{2}{5} \; x \; e^2\,, \qquad
F_{x,\lambda_\BL}^{(1)} = - \frac{\lambda_\BL}{2\pi} \; x \; e^2\,. 
\ee
Notice that the direction of the above forces changes depending on the sign of $e^2$, i.e.~depending on whether the star is oblate or prolate. Notice also that Eq.~\eqref{eq:force-balance-x} states that the centrifugal force, which acts always away from the stellar center, needs to balance the sum of $F_{x,0}^{(1)}$  and $F_{x,\lambda_\BL}^{(1)}$, which from Eq.~\eqref{eq:fx0} is given by
\be
\label{eq:force-sum}
F_{x,0}^{(1)} + F_{x,\lambda_\BL}^{(1)} = - \frac{4\pi+5\lambda_\BL}{10\pi} x e^2\,.
\ee

\begin{figure}[htb]
\begin{center}
\includegraphics[width=8.5cm,clip=true]{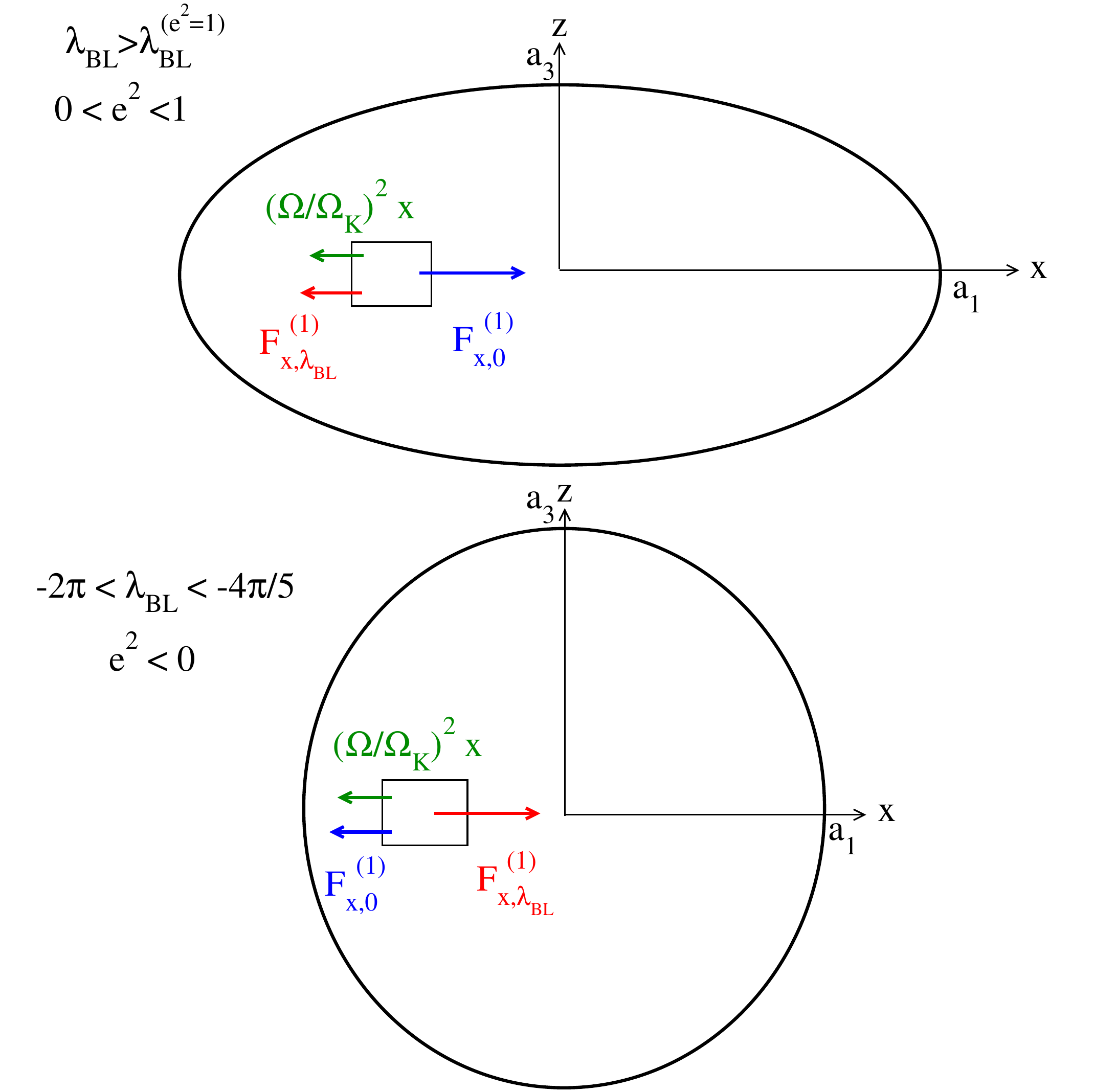}  
\caption{\label{fig:force-balance} (Color online) A schematic picture that shows a free body diagram for a fluid element in an anisotropic star at second order in spin on the equatorial plane.  
$(\Omega/\Omega_\K)^2 x$ is the centrifugal force, while $F_{x,0}^{(1)}$ is the sum of the pressure gradient and gravitational force with isotropic pressure and $F_{x,\lambda_\BL}^{(1)}$ corresponds to its anisotropic correction. The top diagram corresponds to the case with $\lambda_\BL > \lambda_\BL^{(e^2=1)}$ and $0<e^2 <1$. Since $F_{x,\lambda_\BL}^{(1)}$ acts outward, the oblateness of the stellar shape increases as one decreases $\lambda_\BL$. Once the anisotropy parameter reaches the critical value $\lambda_\BL^{(e^2=1)}$ [defined in Eq.~\eqref{eq:lambda-e1}], the star becomes infinitely oblate. One cannot construct an equilibrium configuration with $\lambda_\BL^{(e^2=1)} < \lambda_\BL < -4\pi/5$ within the slow-rotation approximation. If one further decreases $\lambda_\BL$, an equilibrium oblate configuration does not exist anymore and the stellar shape changes to prolate, whose force balance is shown in the bottom diagram for $-2\pi < \lambda_\BL < -4\pi/5$ and $e^2 <0$.
}
\end{center}
\end{figure}

With these expressions at hand, let us now discuss how force balance and the stellar shape change as one varies $\lambda_\BL$, with a fixed $\Omega/\Omega_\K$. Consider first the isotropic case. Setting $\lambda_{\BL} = 0$ in Eq.~\eqref{eq:fx0}, the anisotropic correction to the pressure gradient force vanishes. Since $F_{x,0}^{(1)}$ needs to balance the centrifugal force, from Eqs.~\eqref{eq:force-balance-x} and~\eqref{eq:fx0}, one finds that $e^2 > 0$, which implies the star must be oblate. 

Let us now imagine we were to add a \emph{small}, negative amount of anisotropy ($|\lambda_{\BL}| \ll 1$ and $\lambda_{\BL} < 0$) to an isotropic configuration.  The additional anisotropy force, $F_{x,\lambda_\BL}^{(1)}$, no longer vanishes, and in fact, it must be centrifugal, as depicted on the top of Fig.~\ref{fig:force-balance}, since $e^{2}>0$ for the background, isotropic configuration. In order for balance to be restored, the star must compensate for this additional force, and the only way to do so is for $F_{x,0}^{(1)}$ to increase, which implies $e^{2}$ must also increase and the star becomes more oblate.

\if0%%%%%%%%%%

Indeed, linearizing Eqs.~\eqref{eq:fx0} and~\eqref{eq:fxL}, one finds
%
\ba
\label{eq:fx0-exp}
F_{x,0}^{(1)} &=& - \Omega^2 x \left[1 - \frac{5}{4\pi} \lambda_{\BL} + {\cal{O}}(\lambda_{\BL}^{2}) \right]\,, 
\\
\label{eq:fxL-exp}
F_{x,\lambda_\BL}^{(1)} &=&  -  \frac{5}{4 \pi} \lambda_{\BL} \Omega^{2} x + {\cal{O}}(\lambda_{\BL}^{2})\,.
\ea
%
The isotropic part of the pressure gradient force continues to be balanced by the centrifugal force, but the former acquires an anisotropic correction that is balanced by a new anisotropic pressure force, as depicted on the top of Fig.~\ref{fig:force-balance}. Since $\lambda_{\BL} < 0$, the new anisotropic pressure force is positive, and using this in Eq.~\eqref{eq:fxL}, we see that $e^{2} > 0$ since $\rho>0$ and the star remains oblate, as expected. 

\fi%%%%%%%%%%%%

Imagine now that we increased the magnitude of $\lambda_{\BL}$ further, always with $\lambda_{\BL} < 0$. As mentioned above, $e^2$ is forced to increase as $\lambda_\BL$ becomes more negative to maintain equilibrium, forcing the star to become more and more oblate. Eventually, $\lambda_\BL$ reaches the critical value $\lambda_\BL^{(e^2=1)}$ of Eq.~\eqref{eq:lambda-e1} where $e^2=1$ and the star would seem to become infinitely oblate. Notice, however, that the small eccentricity approximation of Eq.~\eqref{eq:F-small-ecc} becomes invalid near this critical point. If one further decreases $\lambda_\BL$, one enters into a forbidden and unphysical region $(-4\pi/5 < \lambda_\BL < \lambda_\BL^{(e^2=1)})$, where $e^2$ becomes larger than unity, which is an artifact of the slow-rotation approximation.

Consider now decreasing $\lambda_\BL$ even further, such that $-2 \pi < \lambda_\BL < - 4\pi/5$. Equation~\eqref{eq:force-sum} shows that the total pressure and potential gradient force cannot balance the centrifugal force when $\lambda_\BL < - 4\pi/5$ unless $e^{2}$ flips sign, i.e.~unless the star becomes prolate as shown in the bottom panel of Fig.~\ref{fig:force-balance}. Moreover, this equation also shows that as $\lambda_\BL$ decreases the magnitude of the total pressure and potential gradient force increases for a fixed $e^2$ (and $x$). This implies that $|e^2|$ must also decrease to keep the sum of forces balanced against the centrifugal force, making the star less prolate as $\lambda_\BL$ decreases. Thus, the change in sign of the pressure and potential gradient force and its anisotropy correction is responsible for the change in stellar shape. 

Let us conclude with a brief discussion of the stability of strongly-anisotropic, rotating stars by investigating how the fluid elements shift and whether the stellar configuration approaches an equilibrium configuration as one varies $\lambda_\BL$. Notice that since Eq.~\eqref{eq:e2} only determines the \emph{ratio} between $a_1$ and $a_3$, for simplicity, we will consider the $a_3={\rm{const}}$ case. This choice allows us to determine how $e^2$ varies purely from the change in $a_1$.
 
Let us first look at the stability of an oblate, rotating anisotropic star by considering adding a finite amount of negative anisotropy to an isotropic configuration ($\lambda_{\BL} = 0$). The additional force due to anisotropy shifts the fluid elements in the centrifugal direction (see the top of Fig.~\ref{fig:force-balance}). This increases $a_1$, and thus increasing $e^{2}$ and making the star more oblate. As we argued before, the equilibrium configuration becomes more oblate as the amount of anisotropy is increased. Thus, the anisotropy force pushes the star toward the equilibrium configuration, making them stable to perturbations in $\lambda_\BL$.

Let us now study the stability of a prolate, rotating star by considering adding a small amount of negative anisotropy to an equilibrium, prolate configuration with some value of $\lambda_{\BL}$ in $-2 \pi < \lambda_\BL < - 4\pi/5$. The force due to additional anisotropy now shifts the fluid elements in the centripetal direction for a fixed $e^{2}$ (see the top of Fig.~\ref{fig:force-balance}). This decreases $a_{1}$, and thus $|e^{2}|$ increases, making the star even more prolate. However, we argued before that the equilibrium configuration of prolate stars becomes less prolate as anisotropy is increased. Therefore, the anisotropy force pushes the star away from the prolate equilibrium configuration. This suggests that prolate anisotropic stars are unstable to perturbations in $\lambda_\BL$\footnote{Reference~\cite{Glampedakis:2013jya} discusses the instability of prolate anisotropic stars in the context of tidal deformations, though further study is needed to understand how that instability relates to the ones discussed here (if at all).}.

%----------------------------
\subsection{Beyond the Weak-field Limit}
\label{sec:PM}

We here derive an analytic expression for the I-Love relation by perturbing about the weak-field limit in a PM approximation, i.e.~in an expansion about small compactness. We extend~\cite{Chan:2014tva} for isotropic stars to anisotropic stars, starting from the exact solution to the Einstein equations for incompressible, non-rotating, anisotropic stars with the BL model given in Eqs.~\eqref{eq:M-n0}--\eqref{eq:nu-n0}~\cite{1974ApJ...188..657B} as our background.

Let us begin by finding the metric perturbation at linear order in spin. To achieve this, we impose the following ansatz for the metric perturbation and the moment of inertia:
\be
A = \sum_{k=0} A^{(k)} C^k\,,
\ee
where $A$ is either $\omega_{1}$ or $I$, with $\omega_1 (R)$ the $l=1$ mode of the $(t,\phi)$ component of the metric perturbation $\omega (R,\theta)$ in a Legendre decomposition. We substitute this ansatz for $\omega_{1}$, together with Eqs.~\eqref{eq:M-n0}--\eqref{eq:nu-n0}, into the Einstein equations and perturb about $C=0$. We then solve the perturbed Einstein equations order by order in C in the interior region, with regularity imposed at the center. We also substitute the above ansatz for $I$ in the exterior solution and perturb about $C=0$. Finally, we match the perturbed interior and exterior solutions order by order in $C$ at the stellar surface to calculate $I$ within the PM approximation. 

Through this procedure, we find that $\bar I $ is given by
\be
\label{eq:Ibar-PM}
\bar I =  \frac{2}{5} \frac{1}{C^2} \sum_{i=1}^{6} \sum_{j=1}^{i} \left[ 1 + c_{ij}^{(\bar I)}  \left(\frac{\lambda_\BL}{\pi} \right)^j C^i   + \mathcal{O} \left( C^7 \right) \right]\,, 
\ee
where the coefficients $c_{ij}^{(\bar I)}$ are explicitly given in Table~\ref{Table:Ibar}. Notice that the leading term in $C$ does not depend on $\lambda_\BL$, which shows that the $\bar I$--$C$ relation is unaffected by anisotropy in the weak-field limit.

One can derive the PM expression for $\lambda_2$ in a similar manner, namely by perturbing the metric and the Einstein equations, solving the perturbed equations order by order in C, and carrying out a matching calculation at the stellar surface. Doing so, we find
\be
\label{eq:lambdabar-PM}
\bar \lambda_2 =  \frac{2 \pi}{4 \pi + 5 \lambda_\BL} \frac{1}{C^5} \sum_{i=1}^{6} \sum_{j=1}^{i+2} \left\{ 1 + c_{ij}^{(\bar \lambda_2)}  \left(\frac{\lambda_\BL}{\pi} \right)^j \left[ \left( 1 + \frac{5 \lambda_\BL}{4 \pi} \right)^{-1} \frac{C}{4} \right]^i  + \mathcal{O} \left( C^7 \right) \right\}\,, 
\ee
where the coefficients $c_{ij}^{(\bar \lambda_2)}$ are given in Table~\ref{Table:lambdabar}. 

We can now invert Eq.~\eqref{eq:lambdabar-PM} perturbatively (in small $C$ or large $\bar{\lambda}_{2}$) to obtain $C$ as a function of $\bar \lambda_2$. Inserting such an expression into Eq.~\eqref{eq:Ibar-PM}, we find the I-Love relation:
\begin{align}
\label{eq:ILove-PM}
\bar I &= \frac{2}{5} \left[ 2 \left(  1 + \frac{5 \lambda_\BL}{4 \pi} \right) \bar \lambda_2 \right]^{2/5} \sum_{i=1}^{6} \sum_{j=1}^{i+2} \left\{ 1 + c_{ij}^{(\bar I \bar \lambda_2)}  \left(\frac{\lambda_\BL}{\pi} \right)^j \left[ \frac{1}{2^{11}} \left( 1 + \frac{5 \lambda_\BL}{4\pi} \right)^{-6} \frac{1}{\bar \lambda_2} \right]^{i/5} \right. \nn \\
& +  \left. \mathcal{O} \left( \bar \lambda_2^{-7/5} \right) \right\}\,.  
\end{align}
Here, the coefficients $c_{ij}^{(\bar{I}\bar \lambda_2)}$ are given in Table~\ref{Table:ILove}.
Equations~\eqref{eq:Ibar-PM}--\eqref{eq:ILove-PM} with $\lambda_\BL =0$ agree with those found in~\cite{Chan:2014tva} in the isotropic limit.
%

%----------------------------
\subsection{Strong-field Limit}
\label{sec:tidal-strong}

Let us now calculate the tidal deformability $\bar \lambda_2$ for incompressible, anisotropic stars in the strong-field limit for specific choices of $\lambda_\BL$. The strong-field limit here refers to the maximum-compactness limit (not to be confused with the BH limit), i.e.~the limit in which $C \to C_{\max}$. Following the analytic strong-field analysis for isotropic stars of~\cite{damour-nagar}, we introduce the new radial coordinate 
\be
x \equiv \left( 1-2\,{\frac {{\it C}\,{R}^{2}}{R_*^{2}}} \right) ^{\gamma}\,,
\ee
where $\gamma$ is given by Eq.~\eqref{eq:gamma}. Notice that the stellar center corresponds to $x=1$, while the stellar surface corresponds to $x=1/3$ for a maximum compactness configuration [see Eq.~\eqref{eq:max-comp}]. 

With this coordinate choice, the radial pressure $p$ and the metric function $\nu$ in Eqs.~\eqref{eq:p-n0} and~\eqref{eq:nu-n0} simplify to 
\begin{align}
p(x) &= -\frac{3}{4 \pi} {\frac {{\it C}\, \left[\left( 1-2\,{\it C} \right) ^{\gamma} -x
 \right] }{R_*^{2} \left[ 3\, \left( 1-2\,{\it C}
 \right) ^{\gamma}-x \right] }}\,, \\
\nu(x) &=  {\frac {\ln  \left[ 3\, \left( 1-2\,{\it C} \right) ^{\gamma}-x \right] -
\ln  \left( 2 \right) }{\gamma}}
\,.
\end{align}
Moreover, this coordinate choice allows us to decouple the differential equation for the metric perturbation $h_2 (R)$ [that corresponds to the $\ell = 2$ mode of $h(R,\theta)$ in Eq.~\eqref{Eq:metric-slow-rot}] from other metric perturbations, leading to a homogeneous, third-order differential equation.

Before solving for the tidal deformability of anisotropic stars, let us review how this is done analytically in the isotropic case. The third-order differential equation for $h_{2}$ can be integrated once, reducing it to a second-order equation. Reference~\cite{damour-nagar} solved this equation for a maximum compactness configuration ($C_{\max} = 4/9$) both in the interior and in the exterior regions, imposing regularity at the center. The authors then matched the interior solution to the exterior solution at the surface, with a certain jump condition due to the discontinuous density at the surface. Doing so, they found the dimensionless tidal deformability $\bar \lambda_2^{c_{\max}} (\lambda_\BL)$ to be
\be
\label{eq:love-0}
\bar \lambda_2^{c_{\max}}  (0) = \frac{72}{5 (308 - 81 \ln 3)} \simeq 0.0658\,,
\ee
for an isotropic, incompressible star of maximum compactness.
 
One can carry out a similar analysis completely analytically for certain choices of $\lambda_\BL$. For example, when $\lambda_\BL = 2 \pi$, the third-order differential equation for $h_2$ for a maximum compactness configuration is given by
\be
 4 (1 - x)^3 x \frac{d^3 h_2^x}{d x^3} + 6 (1 - x)^2 (1 - 4 x) \frac{d^2 h_2^x}{d x^2} + (19 - 46 x + 27 x^2) \frac{d h_2^x}{d x} - (5 - 3 x) h_2^x = 0\,,
\ee
where $h_2^x (x) \equiv h_2[R(x)]$. Imposing regularity at the center ($x=1$), the solution to the above equation is
\be
h_2^x(x) = -\frac{\sqrt{2} C_h \exp \left(-\sqrt{5} \, \arctanh{\sqrt{x}} \right)}{\sqrt{5(1 -  x)}}\,,
\ee
where $C_h$ is an integration constant, from which one can find
\be
\label{eq:y}
y(R_*) \equiv \frac{h_2'(R_*) R_*}{h_2(R_*)} = \sqrt{15} - 1\,.
\ee
We find that one does not need to worry about the density discontinuity at the surface when $\lambda_\BL = 2 \pi$, because the solution is not discontinuous at the surface for that value of $\lambda_{\BL}$. Hence, one can use Eq.~(23) in~\cite{hinderer-love} with $y$ given by Eq.~\eqref{eq:y} to find the tidal Love number $k_2$. The latter can be turned into the tidal deformability $\lambda_2$ as explained in~\cite{hinderer-love} to find
\be
\label{eq:love-2pi}
\bar \lambda_2^{c_{\max}}  (2 \pi) = \frac{48}{180 + 8 \sqrt{15} - 135 \ln 3} \simeq 0.766\,.
\ee
Notice that the tidal deformability is much larger in the anisotropic case with $\lambda_\BL = 2 \pi$ than in the isotropic one. 

%----------------------------
\subsection{Strongly Anisotropic Limit}
\label{sec:strong-ani}

Let us now analytically investigate the strongly anisotropic limit, $\lambda_\BL \to - 2\pi$, for incompressible anisotropic compact stars of arbitrary compactness. The goal of this subsection is to prove that $\bar I = 4$ in such a limit as $C \to 1/2$, which agrees with the expected result for BHs. To achieve this goal, we analytically construct slowly-rotating anisotropic incompressible compact stars to linear order in spin. 

Let us first discuss the (background) zeroth-order in spin solution. From Eq.~\eqref{eq:p-n0}, one finds that the radial pressure vanishes in the strongly anisotropic limit~\cite{1974ApJ...188..657B,Glampedakis:2013jya}. Taking the limit of $\lambda_\BL \to -2 \pi$ in Eq.~\eqref{eq:nu-n0}, one further finds that 
\be
\nu^{(\inter)} (R) = \frac{3}{2} \ln(1-2 C ) - \frac{1}{2} \ln \left(1 - 2 C \frac{R^2}{R_*^2} \right)\,,
\ee
where the superscript (int) reminds us that this is the solution in the interior region.

Let us now find the interior solution at linear order in spin. Substituting the above background solution into the differential equation for $\omega_1$ (the $\ell=1$ mode of $\omega (R,\theta)$ in Eq.~\eqref{Eq:metric-slow-rot} of~\cite{Yagi:2015hda}), one finds
\be
\frac{d^2 \omega_1^{(\inter)}}{dR^2} + \frac{11 C R^2-4 R_*^2}{R (2 C  R^2 - R_*^2)} \frac{d\omega_1^{(\inter)}}{dR} +  6 C \frac{3 C R^2 - 2 R_*^2}{(2 C R^2 - R_*^2)^2} \omega_1^{(\inter)} = 0\,.
\ee
Imposing regularity at the center, we find an analytic solution for $\omega_1$ in the interior region:
\be
\omega_1^{(\inter)} (R) = C_{\omega_1} (R_*^2 - 2 C R^2)^{3/4} \; {}_2F_1 \left( \frac{3}{2}, \frac{9}{4} ; \frac{5}{2}; 2 C \frac{R^2}{R_*^2} \right)\,,
\ee
where $C_{\omega_1}$ is an integration constant and $_{2}F_{1}(\cdot,\cdot ;\cdot ;\cdot)$ represents hypergeometric functions. 

Matching the above interior solution to the exterior solution, $\omega_1^{\ext} (R) = \Omega (1-2 I/R^3)$, at the stellar surface with the conditions $\omega_1^{(\inter)} (R_*) = \omega_1^{(\ext)} (R_*)$ and $\omega_1^{(\inter)}{}' (R_*) = \omega_1^{(\ext)}{}' (R_*)$, one finds~\cite{Yagi:2015upa}
\be
\label{eq:crit-exp-I}
\bar I (C)  = \frac{1}{2 C^{2}} \frac{\sum_{i=0}^{1} a_{i}(C) \; C^{i}}{\sum_{j=0}^{2} b_{i}(C) \; C^{j}}\,,
\ee
where the coefficients of the numerator are
\begin{align}
a_{0}(C) &= -9 \; {\mbox{$_2$F$_1$}\left(\frac{5}{2},{\frac {13}{4}};\,\frac{7}{2};\,2\,{\it C}\right)} +5 \; {\mbox{$_2$F$_1$}\left(\frac{3}{2},\frac{9}{4};\,\frac{5}{2};\,2\,{\it C}\right)}\,,
\\
a_{1}(C) &= 18 \; {\mbox{$_2$F$_1$}\left(\frac{5}{2},{\frac {13}{4}};\,\frac{7}{2};\,2\,{\it C}\right)}\,,
\end{align}
while the coefficients of the denominator are
\begin{align}
b_{0}(C) &= -5\, {\mbox{$_2$F$_1$}\left(\frac{3}{2},\frac{9}{4};\,\frac{5}{2};\,2\,{\it C}\right)}\,,
\\
b_{1}(C) &= - 9 \; {\mbox{$_2$F$_1$}\left(\frac{5}{2},{\frac {13}{4}};\,\frac{7}{2};\,2\,{\it C}\right)} + 15 \; {\mbox{$_2$F$_1$}\left(\frac{3}{2},\frac{9}{4};\,\frac{5}{2};\,2\,{\it C}\right)}\,,
\\
b_{2}(C) &= 18 \;  {\mbox{$_2$F$_1$}\left(\frac{5}{2},{\frac {13}{4}};\,\frac{7}{2};\,2\,{\it C}\right)}\,.
\end{align}
%
%with $_{2}F_{1}(\cdot,\cdot,\cdot,\cdot)$ hypergeometric functions. 
%
%%
%\bw
%\be
%\label{eq:crit-exp-I}
%\bar I (C)  =  \frac{18\,{\it C}\,
%{\mbox{$_2$F$_1$}(\frac{5}{2},{\frac {13}{4}};\,\frac{7}{2};\,2\,{\it C})}-9\,
%{\mbox{$_2$F$_1$}(\frac{5}{2},{\frac {13}{4}};\,\frac{7}{2};\,2\,{\it C})}+5\,
%{\mbox{$_2$F$_1$}(\frac{3}{2},\frac{9}{4};\,\frac{5}{2};\,2\,{\it C})} } {  2 C^2 \left[ 18\,{{
%\it C}}^{2}
%{\mbox{$_2$F$_1$}(\frac{5}{2},{\frac {13}{4}};\,\frac{7}{2};\,2\,{\it C})}-9\,{\it 
%C}\,{\mbox{$_2$F$_1$}(\frac{5}{2},{\frac {13}{4}};\,\frac{7}{2};\,2\,{\it C})}+15\,{
%\it C}\,{\mbox{$_2$F$_1$}(\frac{3}{2},\frac{9}{4};\,\frac{5}{2};\,2\,{\it C})}-5\,
%{\mbox{$_2$F$_1$}(\frac{3}{2},\frac{9}{4};\,\frac{5}{2};\,2\,{\it C})} \right] }\,. 
%\ee
%\ew
%%
We can now Taylor expand the above expression about $C=C_\BH$. For a slowly-rotating BH, $C_{\BH}  =1/2 + {\cal{O}}(\chi^{2})$, where $\chi$ is the dimensionless spin parameter. With this, one then finds that
\be
\bar I (C)  =  4 - 40 (C-C_\BH) + 224 (C-C_\BH)^2  +  \mathcal{O}\left[  (C-C_\BH)^3, \chi^2 \right]\,.
\ee
In particular, in the $C \to C_{\BH}$ limit, one finds that $\bar I (1/2) = 4$, which agrees with the BH result. Therefore, the above analysis analytically proves that the moment of inertia for strongly anisotropic compact stars approaches the BH limit exactly as the compactness goes to the compactness of a non-spinning BH.

%%%%%%%%%%%%%%%%%%%%%%%%%%%%%%%%%%%
\section{Numerical Results}
\label{sec:BH-limit}

In this section, we investigate numerically how $\bar I$, $\bar \lambda_2$, $\bar Q$ and $\bar S_3$ approach the BH limit as one increases the stellar compactness. We first consider these quantities for anisotropic incompressible stars in GR, and then consider stars in dCS gravity, exploring both the scalar dipole charge and the dCS correction to the moment of inertia as a function of $C$. The results of the previous section are used here to validate our numerical calculations. This section thus shows explicitly how the I-Love-Q relations and the multipole moments approach the BH limit using an equilibrium sequence of strongly anisotropic stars of ever increasing compactness.

%----------------------------------
\subsection{Approaching the BH Limit in GR}
\label{sec:GR}

\begin{figure*}[htb]
\begin{center}
\includegraphics[width=7.5cm,clip=true]{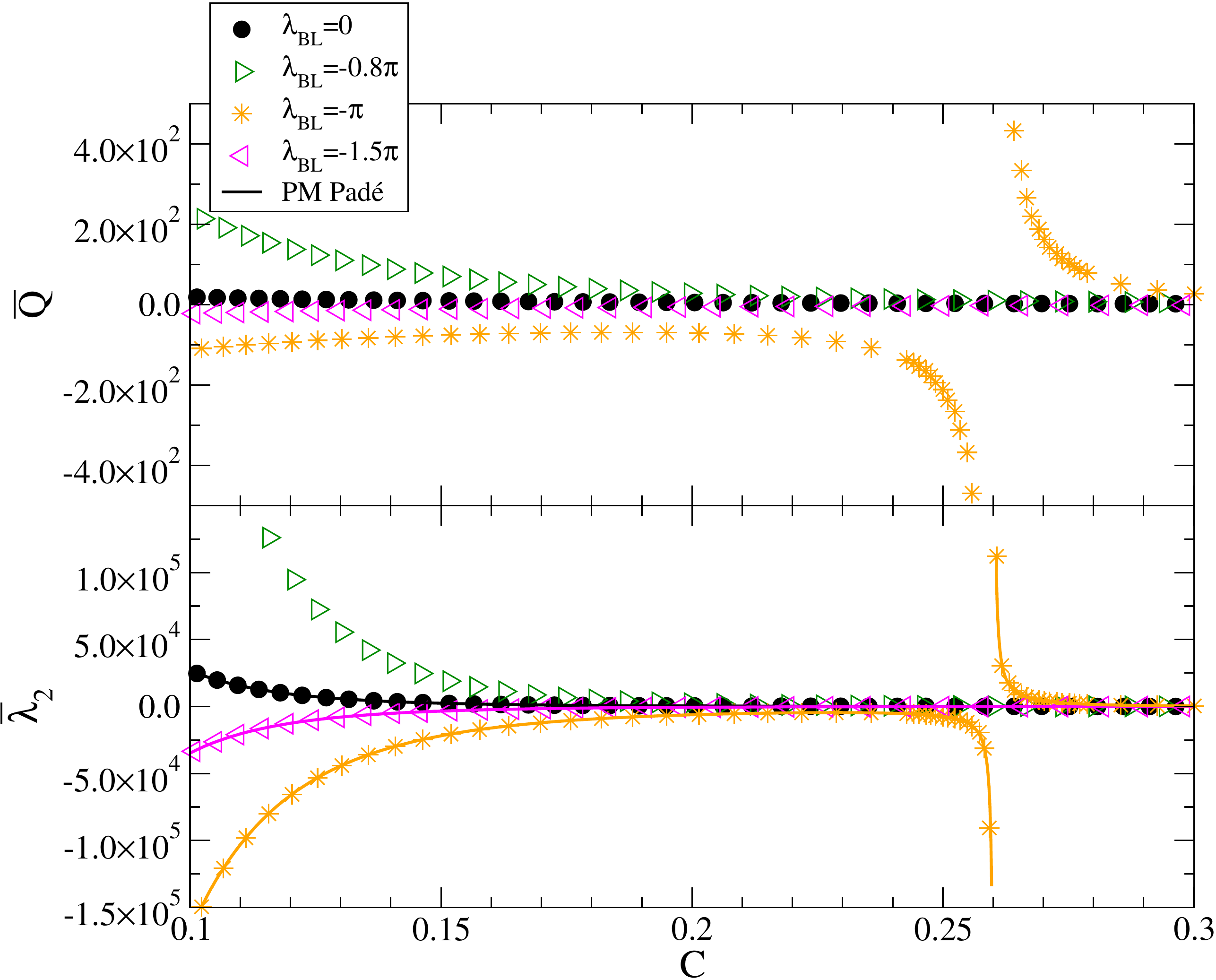}  \quad
\includegraphics[width=7.5cm,clip=true]{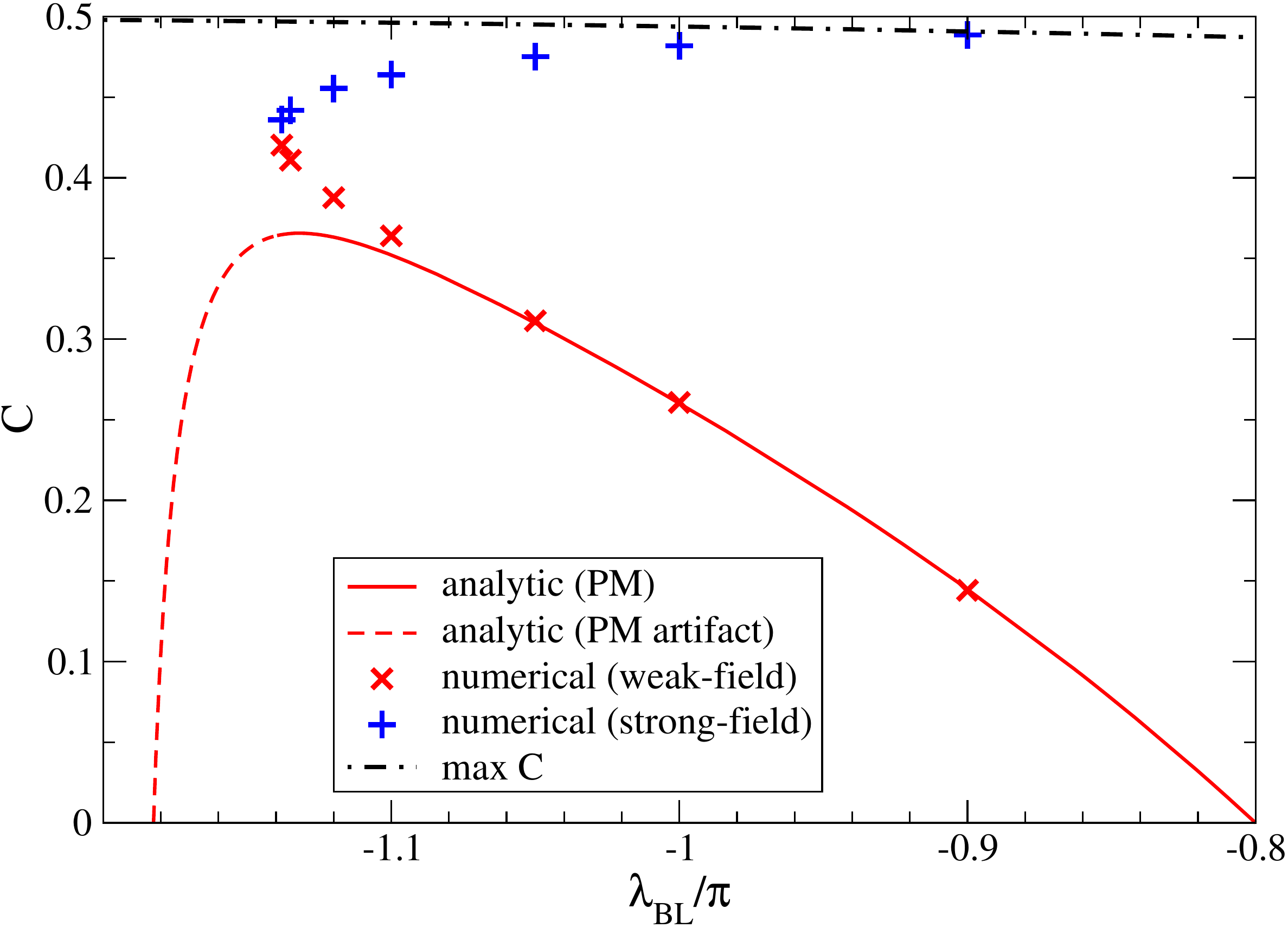}  
\caption{\label{fig:C-dep-n0} (Color online) (Left) Compactness dependence of $\bar Q$ (top) and $\bar \lambda_2$ (bottom) for incompressible stars in the compactness range $C \in (0.1,0.3)$ for various anisotropy parameters. We also show the Pad\'e-resummed, analytic Love-C relation for various values of $\lambda_\BL$ [Eq.~\eqref{eq:33Pade}] with solid curves in the bottom panel.
(Right) Stellar compactness at which the Love-C relation diverges as a function of $\lambda_{\BL}$, as derived from the Pad\'e resummation of the tidal deformability (solid). The dashed red branch ($\lambda_\BL \lesssim -1.14$) is an artifact of the Pad\'e resummation. Points are obtained from numerical calculations in the weak-field (red crosses) and strong-field (blue pluses) regimes. Observe that divergences occur only for $- 1.14 \pi \lesssim \lambda_\BL \leq -0.8 \pi$ in the weak-field regime and $- 1.14 \pi \lesssim \lambda_\BL \lesssim -0.9 \pi$ in the strong-field regime, and the two branches merge as $\lambda_\BL$ becomes more negative. Observe also that the PM approximation becomes invalid for $C \gtrsim 0.35$. We also show the maximum compactness in Fig.~\ref{fig:C-max-BL-n0-noH} as a black dotted-dashed curve, which crosses the blue points at $\lambda_\BL \sim 0.9\pi$.}
\end{center}
\end{figure*}

Let us first investigate how $\bar Q$ and $\bar \lambda_2$ depend on the compactness within the range that is realistic for neutron stars and quark stars. The left panel of Fig.~\ref{fig:C-dep-n0} presents these quantities as a function of compactness within $0.1 \leq C \leq 0.3$. Observe that $\bar Q$ and $\bar \lambda_{2}$ are always positive for incompressible stars with isotropic pressure or with $\lambda_\BL > -0.8 \pi$, where the latter corresponds to the critical value in the weak-field limit~\cite{Glampedakis:2013jya}. On the other hand, when $\lambda_\BL = - \pi$, $\bar Q$ and $\bar{\lambda}_{2}$ are at first negative when $C$ is small, and then they diverge at around $C = 0.26$. For even more anisotropic configurations, such as $\lambda_\BL = -1.5\pi$, $\bar Q$ and $\bar{\lambda}_{2}$ are always negative when $C\leq0.3$ and they do not diverge.

\begin{figure*}[htb]
\begin{center}
\includegraphics[width=7.5cm,clip=true]{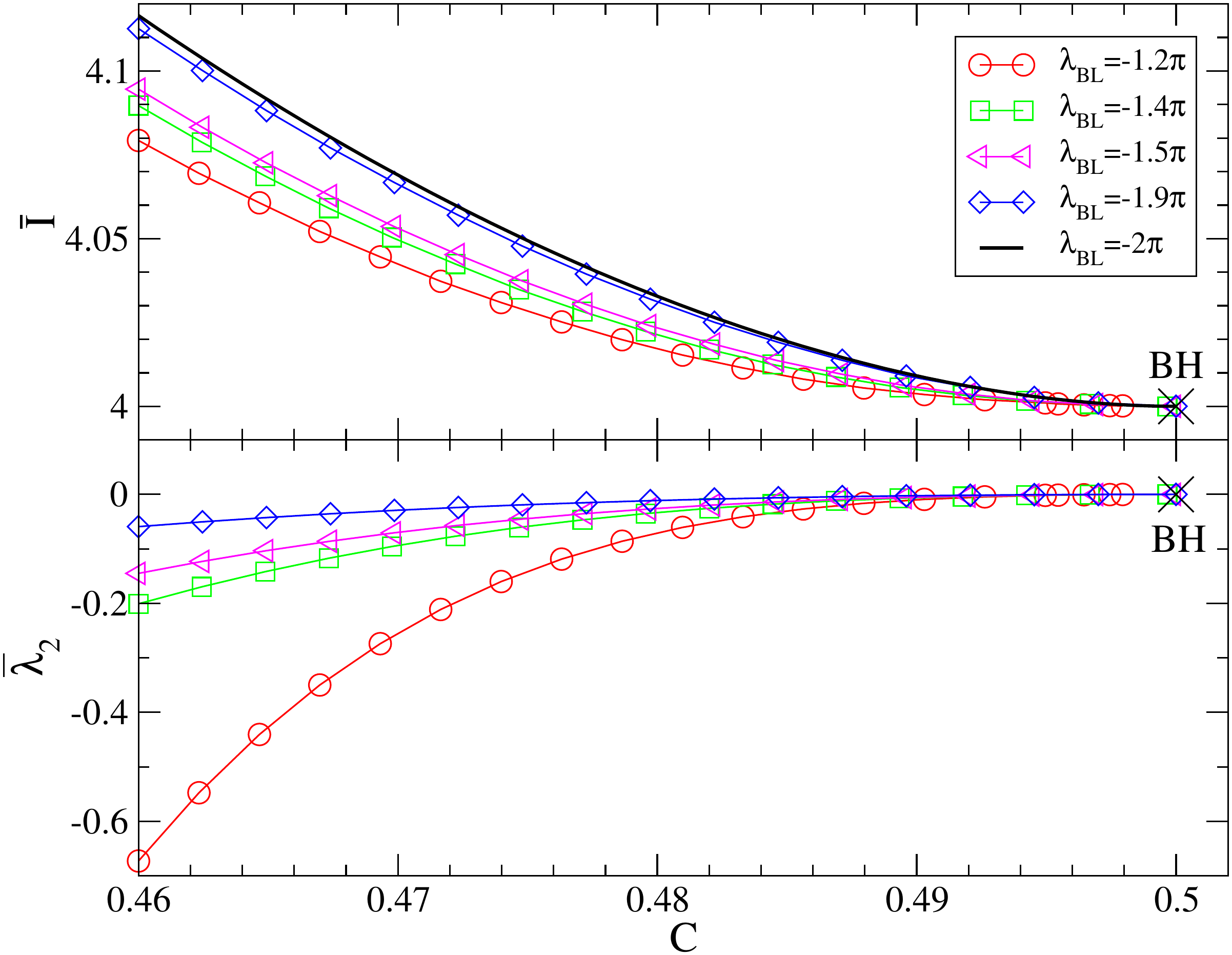}  \quad
\includegraphics[width=7.5cm,clip=true]{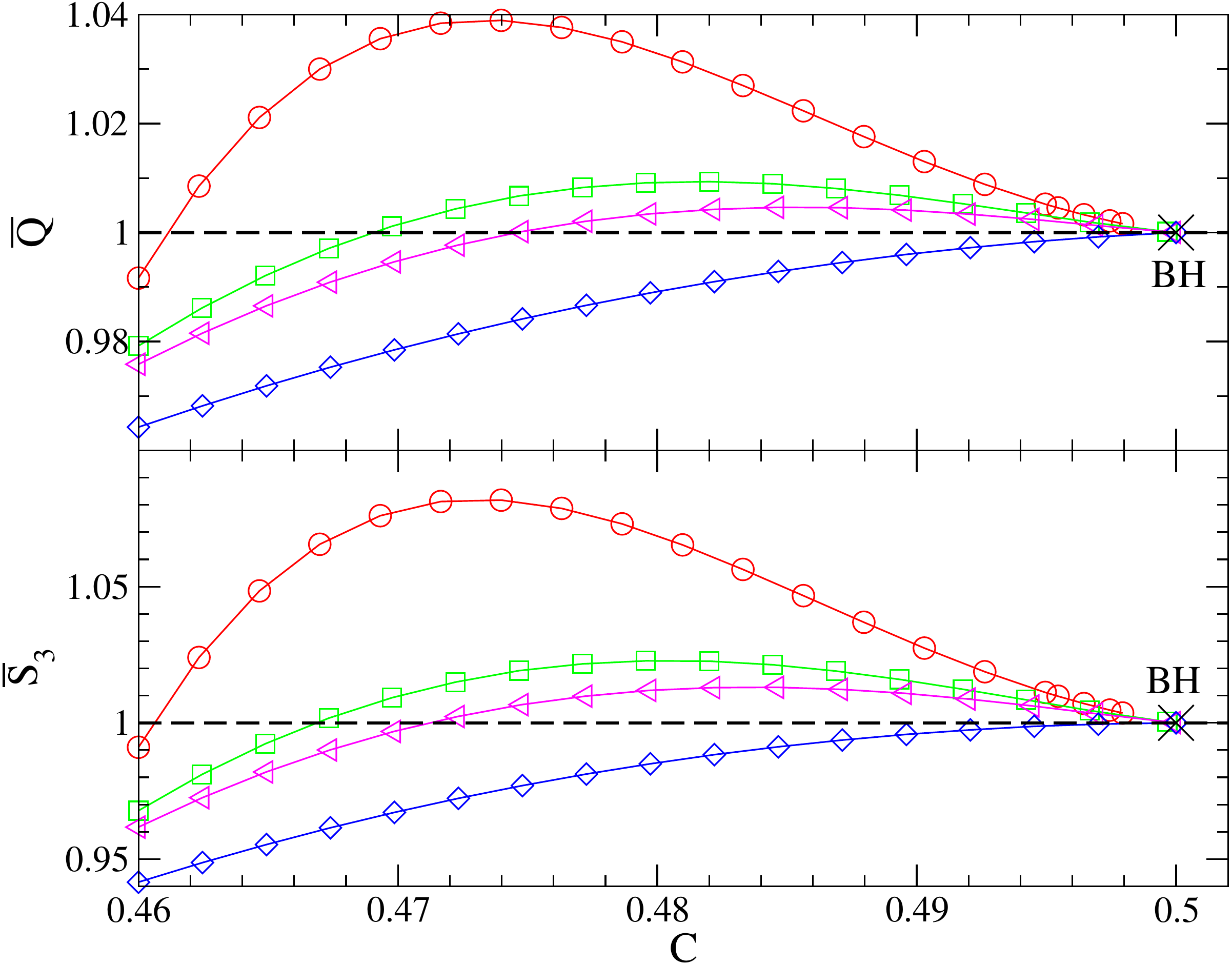}  
\caption{\label{fig:C-dep-polyn0} (Color online) Compactness dependence of $\bar I$ (top left), $\bar Q$ (top right), $\bar \lambda_2$ (bottom left) and $\bar S_3$ (bottom right) for the strongly anisotropic, incompressible stars in the strongly relativistic regime. The solid black curve in the top left panel corresponds to the analytic relation [Eq.~\eqref{eq:crit-exp-I}] for $\lambda_{\BL} = -2\pi$. Dashed lines on the right panel corresponds to the BH value for $\bar Q$ and $\bar S_3$. Observe how each observable approaches the BH result (black cross) as one increases the compactness. 
}
\end{center}
\end{figure*}

Let us find all the values of compactness for which $\bar \lambda_2$ diverges. We can do so analytically through the (3,3)-Pad\'e resummation in Eq.~\eqref{eq:33Pade} of the Taylor series of $\bar \lambda_2$ in Eq.~\eqref{eq:lambdabar-PM}. The bottom, left panel of Fig.~\ref{fig:C-dep-n0} plots this Pad\'e resummation\footnote{We do not show the Pad\'e resummation when $\lambda_\BL = -0.8 \pi$, since then the PM Taylor series diverges.}, which agrees very well with the numerical results, validating the latter. We can also find the location of the divergencies numerically by calculating $\bar{\lambda}_{2}$ as a function of $C$. Doing so, we typically find that the deformability diverges at two values of compactness: a low or weak-field value and a high or strong-field value. The values of compactness for which the tidal deformability diverges are shown in the right panel of Fig.~\ref{fig:C-dep-n0}. Observe that the Pad\'e resummation can only recover the weak-field branch of the divergences, becoming highly inaccurate when $C \gtrsim 0.35$. Observe also that $\bar \lambda_2$ diverges only when $- 1.14 \pi \lesssim \lambda_\BL \leq -0.8 \pi $ for $0 \leq C \lesssim 0.43$ and when $- 1.14 \pi \lesssim \lambda_\BL \leq -0.9 \pi $ for $0.43 \leq C \lesssim C_{\max}$. In particular, these divergences are absent in the BH limit, i.e.~as $\lambda_{\BL} \to -2 \pi$ and $C \to C_\BH$.

What is the physical meaning of these divergences? The right panel of Fig.~\ref{fig:C-dep-n0} shows that there exists a range of $\lambda_{\BL}$ for which the tidal deformability diverges for all values of compactness, and the same would be true of the divergences in the quadrupole moment. This suggests that we can write the location of the divergences as $\lambda_{\BL}^{\rm div} = f_{\bar{Q},\bar{\lambda}_{2}}(C)$, for some functions $f_{\bar{Q},\bar{\lambda}_{2}}(C)$ defined only for $C \in (0,C_{\max})$. In the PM regime, we can asymptotically expand this function about zero compactness 
\be
\label{eq:asy-exp}
f^{\rm wf}_{\bar{Q},\bar{\lambda}_{2}}(C) = -\frac{4 \pi}{5} + \sum_{n=1}^{\infty} a^{(\bar{Q},\bar{\lambda}_{2})}_{n} \; C^{n}\,.
\ee
Similarly, in the strong-field regime, we can expand about $C_{\max}^{(b_{0})}$ with $\lambda_\BL = b_0 \approx -{9 \pi}/{10}$ (where the strong-field branch in blue crosses the maximum compactness curve in black in the right panel of Fig.~\ref{fig:C-dep-n0})
\be
\label{eq:asy-exp-sf}
f^{\rm sf}_{\bar{Q},\bar{\lambda}_{2}}(C) = b_{0} + \sum_{n=1}^{\infty} b^{(\bar{Q},\bar{\lambda}_{2})}_{n} \; \left(C - C_{\max}^{(b_{0})}\right)^{n}\,.
\ee
In both cases we have factored out the weak-field and the strong-field limits. Recall from Sec.~\ref{sec:weak} that in the weak-field limit ($C \to 0$), the divergence in the multipole moments at $\lambda_\BL = -4\pi/5$ was shown to be an artifact of the slow-rotation approximation [see Eq.~\eqref{eq:e2}]. Therefore, the weak-field divergences captured by the asymptotic expansion in Eq.~\eqref{eq:asy-exp} are also due to the slow-rotation approximation in the $\bar{Q}$ case and the small-tide approximation in the $\bar{\lambda}_{2}$ case. Moreover, since $f^{\rm wf}_{\bar{Q},\bar{\lambda}_{2}}$ and $f^{\rm sf}_{\bar{Q},\bar{\lambda}_{2}}$ are two different asymptotic representations of the \emph{same} functions $f_{\bar{Q},\bar{\lambda}_{2}}(C)$, the strong-field divergences captured by the asymptotic expansion in Eq.~\eqref{eq:asy-exp-sf}, and in fact all of the divergences captured by $f_{\bar{Q},\bar{\lambda}_{2}}(C)$, are due to the use of these approximations. Obviously, these approximations become invalid around the region where $\bar Q$ and $\bar \lambda_2$ diverge, since then the metric perturbations become larger than the background and higher-order contributions can no longer be neglected. These divergences would be absent if we had found solutions without the slow-rotation or small-tide approximations.  

Let us now study the compactness dependence of $\bar I$, $\bar \lambda_2$, $\bar Q$ and $\bar S_3$ in the strong-field regime. Figure~\ref{fig:C-dep-polyn0} presents these quantities as a function of the stellar compactness in the range $0.46 \leq C \leq 0.5$. Observe that $\bar I$ ($\bar \lambda_2$) monotonically decreases (increases) to approach the BH limit. Observe also that the analytic $I$--$C$ relation in the strongly anisotropic limit [$\lambda_\BL = -2\pi$, see Eq.~\eqref{eq:crit-exp-I}] validate the numerical results, approximating the latter very closely even when $\lambda_\BL = -1.9\pi$. On the other hand, the behavior of $\bar Q$ and $\bar S_3$ as the BH limit is approached is quite different for some values of $\lambda_{\BL}$. Indeed, these quantities first overshoot the BH values, and then decrease toward the BH limit for $\lambda_\BL = -1.2\pi, -1.4\pi$, and $-1.5\pi$. When $\lambda_\BL = - 1.9 \pi$, $\bar Q$ and $\bar S_3$  both increase monotonically to approach the BH limit. 

\begin{figure}[htb]
\begin{center}
\includegraphics[width=8.5cm,clip=true]{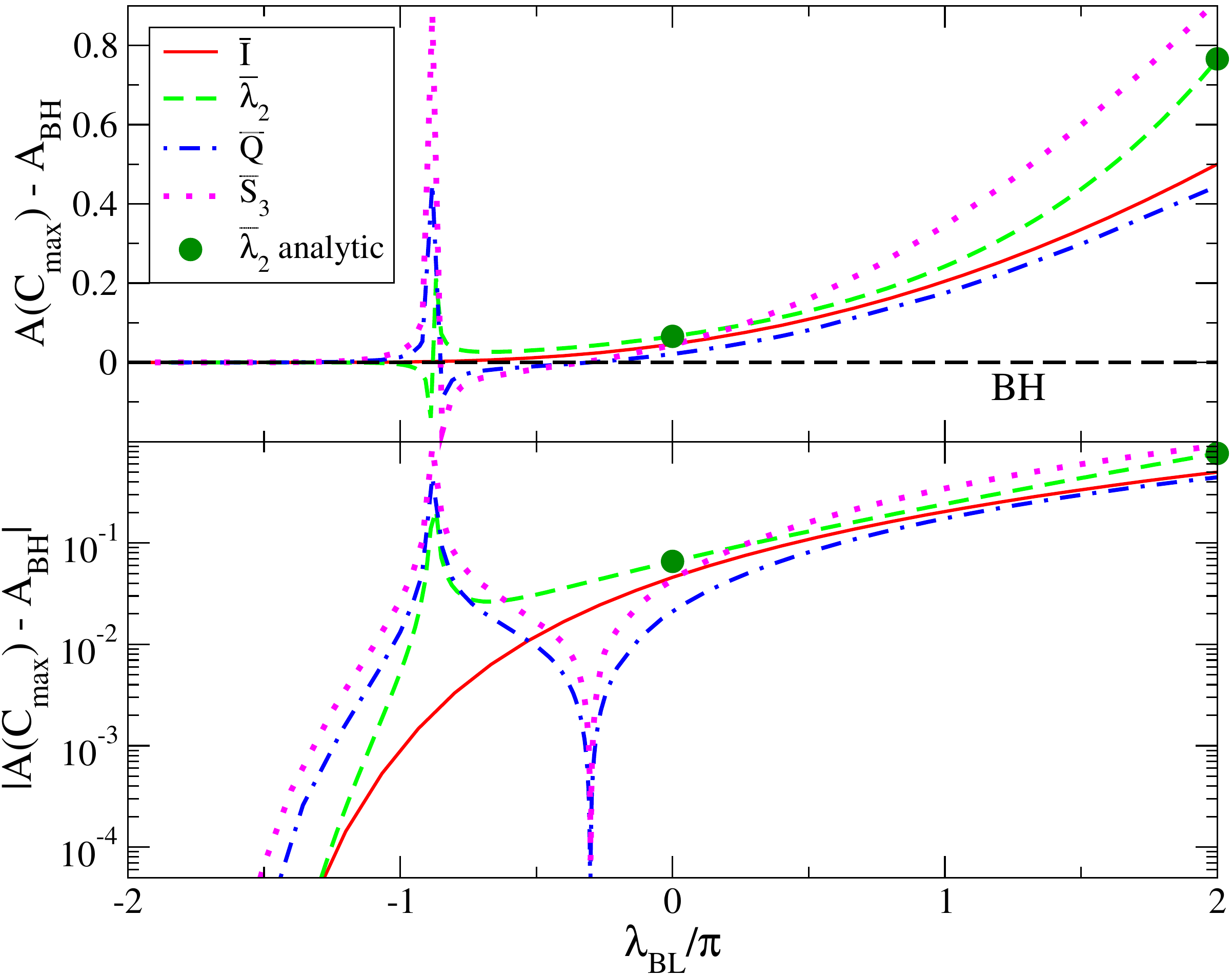}  
\caption{\label{fig:Love-lambda0-Cmax-all} (Color online) (Top) Difference between the properties ($\bar I$, $\bar Q$, $\bar \lambda_2$ and $\bar S_3$) of anisotropic, incompressible stars at their maximum compactness $C_{\max}$ and the properties of a non-rotating BH, plotted as a function of $\lambda_\BL/\pi$. The black dashed horizontal line corresponds to the BH result. Green dots represent analytic values of $\bar \lambda_2$ at $\lambda_\BL = 0, 2 \pi$, given in Eqs.~\eqref{eq:love-0} and~\eqref{eq:love-2pi}. (Bottom) Same as the top panel but the absolute difference in a log plot. Observe how rapidly each observable approaches the BH result as one takes the $\lambda_\BL \to -2 \pi$ limit.
}
\end{center}
\end{figure}
In order to quantify how the properties of anisotropic compact stars approach those of BHs, we take the difference between different observables in the strong-field limit ($C \to C_{\max}$) and the corresponding BH values. Figure~\ref{fig:Love-lambda0-Cmax-all} plots this difference as a function of $\lambda_\BL$, together with the analytic results of Eqs.~\eqref{eq:love-0} and~\eqref{eq:love-2pi} for $\bar \lambda_2 (C_{\max})$ at $\lambda_\BL = 0$ and $2\pi$ (green dots) to validate our numerics. Observe how rapidly each quantity approaches the BH limit as $\lambda_\BL \to -2\pi$, or equivalently, as $C_{\max} \to 1/2$, since recall that $C_{\max} \to 1/2$ only when $\lambda_\BL \to -2\pi$ [Fig.~\ref{fig:C-max-BL-n0-noH}]. Observe also that $\bar \lambda_2 (C_{\max})$, $\bar Q (C_{\max})$ and $\bar S_3 (C_{\max})$ diverge at $\lambda_\BL \sim -0.9\pi$; this value of $\lambda_{\BL}$ corresponds to that where the strong-field branch in blue crosses the maximum compactness curve in black in the right panel of Fig.~\ref{fig:C-dep-n0}, and, as we discussed before, they are artifacts of the slow-rotation and small-tide approximations.

\begin{figure}[htb]
\begin{center}
\includegraphics[width=8.5cm,clip=true]{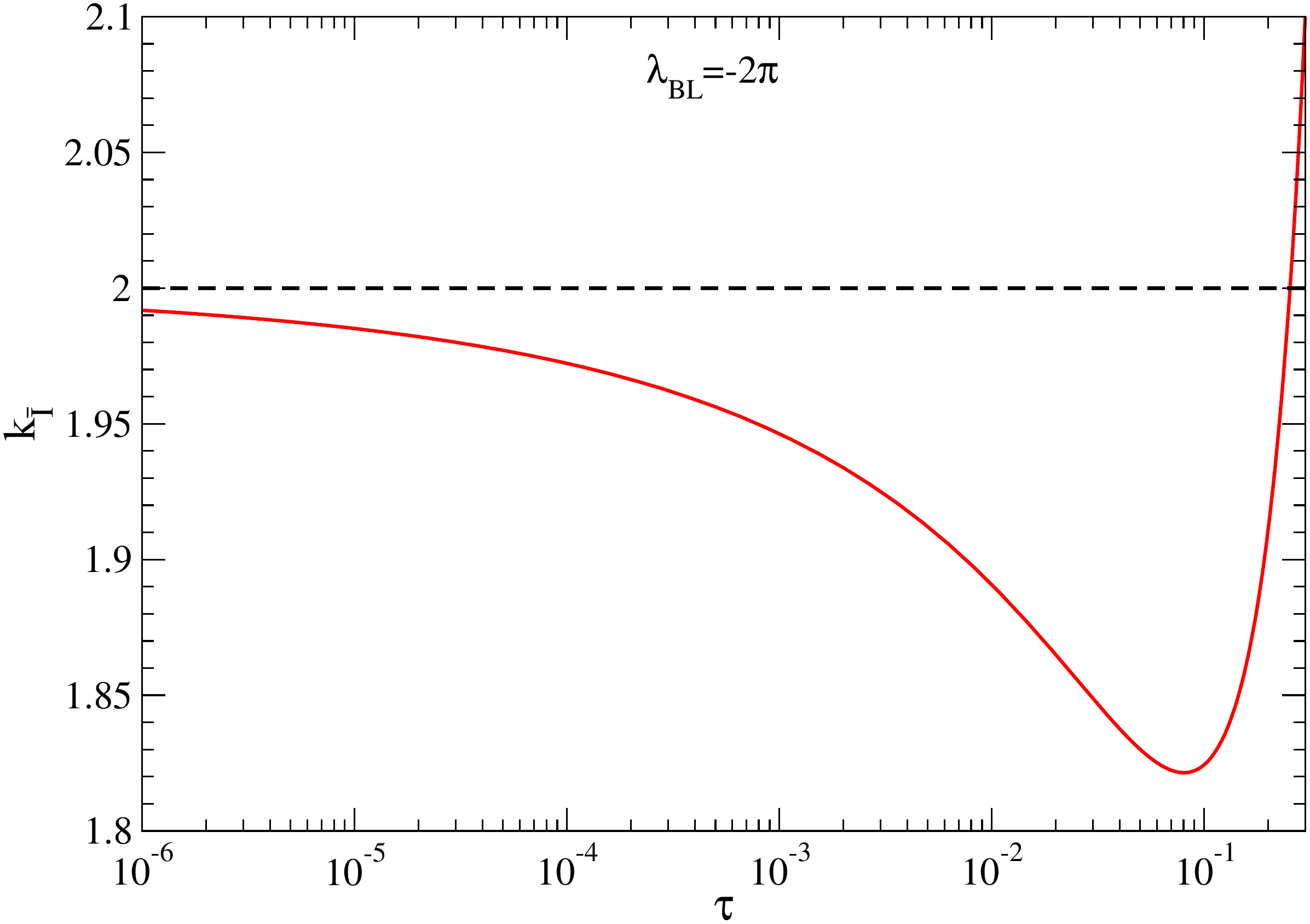}  
\caption{\label{fig:crit-exp-I} (Color online) Scaling exponent of $\bar I$ as a function of $\tau$ with $\lambda_\BL = -2 \pi$, obtained analytically using Eq.~\eqref{eq:crit-exp-I}. Observe that the exponent approaches 2 (black dashed) as one approaches the BH limit of $\tau \to 0$.
}
\end{center}
\end{figure}

Let us now investigate how fast $\bar I$ approaches the BH limit as one takes the $C \to 1/2$ limit. In order to quantify this, we define the scaling exponent of $\bar I$ as
\be
k_{\bar I} \equiv \frac{d \ln \Delta \bar I}{d \ln \tau}\,,
\ee
where 
\be
\tau \equiv \frac{C_\BH - C}{C_\BH}\,, \quad \Delta \bar I \equiv \frac{\bar I - \bar I_\BH}{\bar I_\BH}\,.
\ee
One can explicitly calculate $k_{\bar I}$ with $\lambda_\BL = -2\pi$ from Eq.~\eqref{eq:crit-exp-I}, which we show in Fig.~\ref{fig:crit-exp-I}. Observe that $k_{\bar I}$ approaches 2 as the BH limit is approached ($\tau \to 0$). In fact, one can expand $k_{\bar I}$ as calculated from Eq.~\eqref{eq:crit-exp-I} around $\tau = 0$ to find analytically that $k_{\bar I} = 2 + \mathcal{O}(\tau^{1/4})$. 
In~\cite{Yagi:2015hda}, we also looked at the scaling exponents of $\bar Q$ and $\bar S_3$ and found that they roughly approach the BH limit linearly and quadratically respectively. 

Let us finally look at the interrelations among these observables, i.e.~the I-Love-Q relations. Figure~\ref{fig:univ-polyn0} already presented the I-Love and Q-Love relations for an equilibrium sequence of anisotropic incompressible stars with increasing compactness (indicated by the arrows) for various choices of $\lambda_\BL$. Notice that we plot the absolute value of $\bar \lambda_2$ and $\bar Q$, since these quantities can be both positive and negative. Observe that the anisotropic I-Love relation is qualitatively very similar to the isotropic one, with both $\bar I$ and $|\bar \lambda_2|$ monotonically decreasing as one increases the compactness, irrespective of $\lambda_\BL$.  On the other hand, the strongly anisotropic Q-Love relation is quite different from the isotropic one, with $|\bar Q|$ decreasing to zero at $|\bar \lambda_2| \sim 10$, but then starting to increase again towards the BH limit. This behavior is due to $\bar Q$ changing sign at this $|\bar \lambda_2|$ when $\lambda_\BL = -1.5\pi, -1.9\pi$. In the top panel of Fig.~\ref{fig:univ-polyn0}, we also show the analytic, (3,3)-Pad\'e approximation to the I-Love relation of Sec.~\ref{sec:PM}, which validates our numerical results, deviating from it only in the strongly relativistic regime where the PM approximation loses accuracy.

%----------------------------------
\subsection{Approaching the BH Limit in dCS Gravity}

Let us now investigate whether the stellar quantities approach the BH values as one increases the compactness even in theories other than GR, taking dCS gravity~\cite{jackiw,Smith:2007jm,CSreview} as an example. This theory is motivated from the theoretical requirement to cancel anomalies in heterotic string theory~\cite{polchinski2}, the inclusion of matter when scalarizing the Barbero-Immirzi parameter in loop quantum gravity~\cite{alexandergates,taveras,calcagni} and from effective theories of inflation~\cite{Weinberg:2008hq}. 

The dCS action modifies the Einstein-Hilbert action by adding a (pseudo-) scalar field with a canonical kinetic term, a $\vartheta$-dependent potential and an interaction term in the form of the product of the Pontryagin density and the scalar field. DCS gravity breaks parity at the level of the field equations, in the sense that it introduces modifications to GR only in spacetimes that break spherical symmetry, like that of rotating compact stars. The strength of dCS modifications are proportional to its coupling parameter $\alpha$, which multiplies the interaction term. Since higher curvature corrections to the action are currently unknown, we treat the theory as an effective field theory and keep terms up to leading order in the coupling constant [$\mathcal{O}(\alpha)$ in the scalar field and $\mathcal{O}(\alpha^2)$ in the metric]. This avoids problems with the initial value formulation that arise when one insists in treating the theory as exact~\cite{Delsate:2014hba}.

In this paper, we focus on how the dimensionless scalar dipole charge $\bar \mu$ and the dCS correction to $\bar I$ approach the BH limit in the strongly relativistic regime. Following~\cite{Yagi:2015hda}, we construct a slowly-rotating anisotropic incompressible star to linear order in spin in dCS gravity with a vanishing scalar field potential. We extract $\bar \mu$ from the asymptotic behavior of the scalar field at spatial infinity~\cite{Yagi:2013mbt}:
\be
\label{eq:dipole}
\vartheta (R,\theta) = \frac{5}{8}\alpha \bar \mu C^3 \chi \frac{\cos\theta}{R^2} + \mathcal{O} \left( \frac{M_*^3}{R^3} \right)\,.
\ee
Imposing regularity at the horizon, one finds the BH limit of the dimensionless scalar charge $\bar \mu_\BH = 8$~\cite{Yagi:2013mbt}. One obtains the dCS correction to $\bar I$ by looking at the asymptotic behavior of the $(t,\phi)$ component of the metric. We introduce $\delta \bar I$ as~\cite{Yagi:2015hda} 
\be
\label{eq:delta-Ibar}
\delta \bar I = \frac{M_*^4}{\xi_\CS} \frac{\bar I^\CS}{\bar I^\GR}\,, 
\ee
where $\bar I^\GR$ and  $\bar I^\CS$ are the GR and dCS contributions to $\bar I$, 
while $\xi_\CS \equiv 16 \pi \alpha^2$. The moment of inertia for a BH in dCS gravity is given by
\be
\bar I_\BH = \frac{S_{1,\BH}}{\Omega_\BH M_\BH^3}\,,
\ee
where $M_\BH$ and $\Omega_\BH$ correspond to the BH mass and the BH angular velocity at the horizon respectively. The latter is given by
\be
\label{eq:Omega-BH}
\Omega_\BH = \Omega_{\Kerr} \left( 1 - \frac{709}{7168} \frac{\xi_\CS}{M_\BH^4} \right)\,,
\ee
where $\Omega_\Kerr$ is the BH angular velocity for the Kerr solution. From Eqs.~\eqref{eq:delta-Ibar}--\eqref{eq:Omega-BH}, one finds that the dCS corrections to the moment of inertia for a BH is
\be
\delta \bar I_\BH = \frac{709}{1792} \approx 0.396\,.
\ee

Figure~\ref{fig:I-Love-Shen-error} shows $\bar \mu$ and $\delta \bar I$ as a function of $C$ in dCS gravity for various values of $\lambda_\BL$. The BH limit in each panel is shown as a black cross. Observe that, unlike in the case in GR, $\bar \mu$ and $\delta \bar I$ do not approach the BH limit, but instead approach a different point. In order to confirm these numerical calculations, we derived analytic, (2,2)-Pad\'e resummed relations between $\bar \mu$ and $C$ within the PM approximation, as given by Eq.~\eqref{eq:22Pade} (solid curves in the top panel of Fig.~\ref{fig:I-Love-Shen-error}). Observe that such analytic relations also show the feature that $\bar \mu$ does not approach the BH limit in the limit $C \to 1/2$. Since $\delta \bar I$ depends on $\bar \mu$, one would not expect $\delta \bar I$ to approach the BH limit either, given that $\bar \mu$ does not.

\begin{table}
\caption{\label{table:exponent} Scaling exponents for $\delta \bar I$ as a function of $\tau$ for incompressible anisotropic stars in dCS gravity.} 
\begin{indented}
\lineup
\item[]\begin{tabular}{c|ccc}
\br                              
$\lambda_\BL$ &  \multicolumn{1}{c}{$-1.2\pi$}
&  \multicolumn{1}{c}{$-1.5\pi$} &  \multicolumn{1}{c}{$-1.9\pi$}   \\
\hline
 $k_{\delta \bar{I}}$  & $-0.256$ & $-0.311$  & $-0.355$ \\
\br
\end{tabular}
\end{indented}
\end{table}

Following~\ref{sec:GR}, let us finally study how the relation between $\delta \bar I$ and $C$ in dCS gravity behaves near the $C = 1/2$ critical point. We estimate the scaling exponent $k_{\delta \bar I}$ of such a relation for various values of $\lambda_\BL$, which we show in Table~\ref{table:exponent}. Observe that the exponents are negative, unlike in the relation between $\bar I$ and $C$ in GR. This means that $\delta \bar I$ diverges in the limit $\tau \to 0$ (or $C \to 1/2$), which is consistent with the bottom panel of Fig.~\ref{fig:I-Love-Shen-error}. The BH values for $\bar \mu$ and $\delta \bar I$ are obtained by imposing regularity at the horizon. Since $\bar \mu$ and $\delta \bar I$ for an anisotropic compact star do not approach the BH values as one takes $C \to 1/2$, such quantities are not regular at the surface and diverge in the limit. In this sense, such a solution is unphysical and it is also indicative of the breakdown of the small coupling approximation. This failure to arrive at the BH limit suggests a failure in the description of the BH limit as an equilibrium sequence of anisotropic compact stars in dCS gravity.

%%%%%%%%%%%%%%%%%%%%%%%%%%%%%%%%%%%
\section{Future Directions}
\label{sec:future}

We have studied how the I-Love-Q relations for compact stars approach the BH limit as one increases the stellar compactness  by considering an equilibrium sequence of slowly-rotating/tidally-deformed, incompressible stars with anisotropic pressure. We found that this sequence approaches the BH limit in a nontrivial way, similar to the way the no-hair like relations approach the BH limit~\cite{Yagi:2015upa}. We have also calculated the scalar dipole charge and moment of inertia for incompressible, anisotropic stars in dCS gravity. We found that, unlike in GR, these quantities do not approach the BH limit as one increases the stellar compactness, and in fact, the metric diverges in this limit. These findings suggest that whether one can use a sequence of anisotropic compact stars as a toy model to probe how the I-Love-Q relations approach the BH limit depends on the underlying gravitational theory.

We have also carried out analytic calculations in various limits to validate and elucidate our numerical results. In the strongly anisotropic limit, we derived the moment of inertia as a function of the compactness and proved analytically that it reaches the BH value in the BH limit. In the weak-field limit, we constructed incompressible, anisotropic spheroids with arbitrary rotation that reduce to Maclaurin spheroids in the isotropic limit. We used such spheroids to explain why strongly anisotropic, rotating stars become prolate in the weak-field limit. We also derived analytic PM expressions for the moment of inertia and tidal deformability, which accurately reproduce the numerical results for $C \lesssim 0.35$.

Our analysis can be extended in various directions. A natural extension would be to study how higher-order multipole moments approach the BH limit. Instead of constructing slowly-rotating solutions, one can construct rapidly-rotating ones using e.g.~the RNS code~\cite{stergioulas_friedman1995} and extract higher order multipole moments. One can also calculate higher-order ($\ell \geq 3$) tidal deformabilities for anisotropic stars by extending the isotropic analysis of~\cite{damour-nagar,binnington-poisson,Yagi:2013sva} and study how they approach the BH limit. One should easily be able to apply some of the analytic techniques presented here to higher-order tidal deformabilities, like the PM analysis in Sec.~\ref{sec:PM} and the calculation in the strong-field limit in Sec.~\ref{sec:tidal-strong}.

Another avenue for future work includes performing a stability analysis of Maclaurin-like spheroids for anisotropic stars. Figure~\ref{fig:Omega2-e2} shows that multiple values of $e^2$ are allowed for a given value of $\Omega/\Omega_\K$ for a fixed value of $\lambda_\BL$. One can study whether one of these branches is unstable to perturbations, as we suggested in Sec.~\ref{sec:shape_analy} (see also~\cite{Glampedakis:2013jya}). One can also construct Jacobi-like ellipsoids by relaxing axisymmetry. Then, one can again carry out a stability analysis to see whether Maclaurin-like or Jacobi-like spheroids are energetically favored.

Yet, another natural extension is to study universal relations for anisotropic stars in non-GR theories other than dCS gravity. For example, Silva \textit{et al}.~\cite{Silva:2014fca} constructed slowly-rotating, anisotropic neutron star solutions to linear order in spin in scalar-tensor theories~\cite{Damour:1992we,Damour:1993hw}. One could repeat their analysis in the strongly anisotropic regime, extract the scalar monopole charge and the moment of inertia, and study whether these quantities approach the BH limit as one increases the stellar compactness.

In this paper, we followed~\cite{Chan:2014tva} and constructed Pad\'e resummed expressions for $\bar I$ and $\bar \lambda_2$. One can extend such an analysis to $\bar Q$ and $\bar S_3$. Our preliminary results show that the (3,3)-Pad\'e approximant for $\bar Q$ cannot capture our numerical results as well as the Pad\'e resummation of the PM expansion of $\bar I$ and $\bar \lambda_2$ even in the isotropic case. One may need to include higher-order terms in $C$ and construct higher-order approximants, or alternatively, one may need to use other resummation methods or consider other expansions beyond the weak-field one.

We considered anisotropic stars as a toy model, whose compactness can reach the BH value in the strongly anisotropic limit. An important future task would be to consider a more realistic situation: the gravitational collapse of compact stars into BHs. One could then try to dynamically monitor the I-Love-Q and no-hair like relations in such time-dependent situations to see how they compare to the results found in this paper and in~\cite{Yagi:2015upa}. Since the Geroch-Hansen multipole moments~\cite{geroch,hansen} used here and in~\cite{Yagi:2015upa} are only defined for stationary spacetimes, one may first need to develop a generalization that is valid in dynamical situations, and yet reduces to the Geroch-Hansen moments in stationarity. One would also need to consider how to extract such generalized moments from numerical gravitational collapse calculations.

A final avenue for future work could be the study of a possible connection between transitions from compact stars to BHs and second-order phase transitions in condensed matter physics~\cite{1992smpt.book.....Y}. In~\cite{Yagi:2015upa}, we showed that the scaling exponents for the moment of inertia (or the current dipole moment), the mass quadrupole moment and the current octupole moment are EoS universal to $\mathcal{O}(10\%)$ for isotropic stars. Such a feature may have some analog with the universality of critical exponents in second-order phase transitions. Using the anti-de Sitter (AdS)/conformal field theory (CFT) correspondence, Refs.~\cite{deBoer:2009wk,Arsiwalla:2010bt} showed that transitions from non-rotating compact stars to BHs in AdS spacetimes correspond to phase transitions from high density, baryonic states to thermal quark-gluon plasma states on the CFT side. 

Having said this, whether this analogy can be firmly established remains to be seen. This is because the exponents found in~\cite{Yagi:2015upa} are positive, while those in second-order phase transitions are negative. Moreover, nobody has yet shown that scale invariance exists in the collapse of realistic compact objects to BHs\footnote{However, see e.g.~\cite{Choptuik:1992jv,Gundlach:2007gc} for scale invariance, universality and critical phenomena arising in the context of critical gravitational collapse of scalar fields.}. If this were the case, one would then have to uncover what the analogy for the correlation length in the gravity sector is. More detailed investigations would thus be necessary to further elucidate this intriguing line of study. 

%%%%%%%%%%%%%%%%%%%%%%%%%%%%%%%%%%%%
%\acknowledgments
\section*{Acknowledgments}

We would like to thank Steven Gubser, Jim Lattimer, Bennett Link and Anton B. Vorontsov for useful comments, suggestions and advice. 
K.Y.~acknowledges support from JSPS Postdoctoral Fellowships for Research Abroad and NSF grant PHY-1305682.
N.Y. acknowledges support from NSF CAREER Grant PHY-1250636. 
Some calculations used the computer algebra-systems MAPLE, in combination with the GRTensorII package~\cite{grtensor}.

%%%%%%%%%%%%%%%%%%%%%%
\appendix

%-----------------------------
\section{Tortoise Coordinates for Anisotropic Compact Stars}
\label{sec:causal}

In this appendix, we derive the radial tortoise coordinate for non-rotating, incompressible and anisotropic stars with $\lambda_\BL = -2\pi$, which we use in Sec.~\ref{sec:ani-model} to study their causal structure (see Fig.~\ref{fig:causal}). We achieve this by transforming our coordinate system to Eddington-Finkelstein coordinates. The causal structure in the exterior region of an anisotropic compact star is the same as that of a Schwarzschild BH. One introduces the null coordinate $v = t + r_*^\ext$, where $ r_*^\ext$ is the radial tortoise coordinate in the exterior region. We change coordinates from $t$ to $v$ in the metric and impose that $g_{vR}= - g_{tt} (d r_*^\ext/dR) = 1$ to find 
\be
\label{eq:tortoise-ext}
\frac{r_*^\ext}{R_*} = -\frac{1}{R_*} \int \frac{dR}{g_{tt}} =  \frac{R}{R_*} + 2 C \ln \left( \frac{1}{2C}\frac{R}{R_*} - 1 \right)\,. 
\ee

Regarding the interior region, one introduces another null coordinate $v = t + r_*^\inter$, where $ r_*^\inter$ is the radial tortoise coordinate in the interior region. In this case, one needs to transform not only the coordinate $t$ to $v$ but also the coordinate $\phi$ to $\psi$ via $\psi = \phi + \bar{r}$, where $\bar{r}$ is a function of $R$ and $\theta$. Such a coordinate transformation is similar to that from the Boyer-Lindquist coordinates to Eddington-Finkelstein coordinates in the Kerr metric. Imposing $g_{vR}=1$ (and $g_{RR}=0$ to further determine $\bar r$), one finds 
\begin{align}
\frac{r_*^\inter}{R_*} &= 1 - \frac{1}{2(1 - 2 C)} +2 \text{C} \log \left(\frac{1}{2 \text{C}}-1\right) + \frac{1}{2 \sqrt{2 C} (1-2 \text{C})^{3/2}} \left[  \frac{\sqrt{2C} R }{R_*} \sqrt{1-2 \text{C} \frac{R^2}{R_*^2}}   \right. \nn \\ 
& \times \left.  \sin ^{-1}\left(\sqrt{2 C} \frac{R}{R_*} \right) - \sin ^{-1}\left(\sqrt{2 C} \right)\right]\,,
\label{eq:tortoise-int}
\end{align}
where we used Eq.~\eqref{eq:nu-n0} for the $(t,t)$ component of the metric and determined the integration constant such that $r_*^\inter(R_*) = r_*^\ext(R_*)$.

\if0%%%%%%%%%%%%%

Let us now investigate how the slope of the outgoing null geodesics change with $C$. The slope of red curves in the figure for anisotropic compact stars with $\lambda_\BL = -2\pi$ can be calculated as 
%
\be
\frac{dT}{dR} = \frac{R + 2 R_* C}{R - 2 R_* C }\,.
\ee
%
Therefore, such a slope diverges at the surface with a star of $C=1/2$ with the rate $dT/dR \propto (1-2C)^{-1}$. 
The spin correction to such a slope at second order is proportional to $(1-2C)^{-2}$ with the proportional constant depending on the internal structure via the quadrupole moment and the spin correction to the mass. Notice that such a spin correction to the slope diverges faster than the non-rotating part, which suggests that the behavior of the slope at the surface with a compactness close to 1/2 can be different for a non-rotating and rotating configuration, with even an infinitesimal spin for the latter. Such a situation is also present in the slowly-rotating Kerr BH case.
 
\fi%%%%%%%%%%%%%%% 

%%%%%%%%%%%%%%%%%%%%%%%%%%%%%%%%%%%%
\section{Tables and Pad\'e Approximants for the PM Analysis}

%
%{\renewcommand{\arraystretch}{1.2}
\fulltable{\label{Table:Ibar} Coefficients $c^{(\bar I)}_{ij}$ in Eq.~\eqref{eq:Ibar-PM} for $\bar I$ as a function of the stellar compactness $C$ within the PM approximation.
\vspace{3mm}}
%\begin{table*}
%\begin{centering}
%\begin{tabular}{ccccccccccccccccccccccccccc}
\hline
\hline
\noalign{\smallskip}
\multicolumn{1}{c}{$c^{(\bar I)}_{1,0}$}
& \multicolumn{1}{c}{$c^{(\bar I)}_{1,1}$} 
& \multicolumn{1}{c}{$c^{(\bar I)}_{2,0}$} 
& \multicolumn{1}{c}{$c^{(\bar I)}_{2,1}$} 
& \multicolumn{1}{c}{$c^{(\bar I)}_{2,2}$} 
& \multicolumn{1}{c}{$c^{(\bar I)}_{3,0}$} 
& \multicolumn{1}{c}{$c^{(\bar I)}_{3,1}$} 
& \multicolumn{1}{c}{$c^{(\bar I)}_{3,2}$} 
& \multicolumn{1}{c}{$c^{(\bar I)}_{3,3}$} 
\\
\hline
\noalign{\smallskip}
\multicolumn{1}{c}{$\frac{6}{7}$} & 
\multicolumn{1}{c}{$-{\frac {3}{28}}$} & 
\multicolumn{1}{c}{${\frac {106}{105}}$} &
\multicolumn{1}{c}{$-{\frac {5}{28}}$} & 
\multicolumn{1}{c}{$-\frac{1}{42}$} & 
\multicolumn{1}{c}{${\frac {316}{231}}$} & 
\multicolumn{1}{c}{$-{\frac {298}{1155}}$} & 
\multicolumn{1}{c}{$-{\frac {155}{1848}}$} & 
\multicolumn{1}{c}{$-{\frac {1}{168}}$} \\
\noalign{\smallskip}
\multicolumn{1}{c}{$c^{(\bar I)}_{4,0}$}
& \multicolumn{1}{c}{$c^{(\bar I)}_{4,1}$} 
& \multicolumn{1}{c}{$c^{(\bar I)}_{4,2}$} 
& \multicolumn{1}{c}{$c^{(\bar I)}_{4,3}$} 
& \multicolumn{1}{c}{$c^{(\bar I)}_{4,4}$} 
& \multicolumn{1}{c}{$c^{(\bar I)}_{5,0}$} 
& \multicolumn{1}{c}{$c^{(\bar I)}_{5,1}$} 
& \multicolumn{1}{c}{$c^{(\bar I)}_{5,2}$} 
& \multicolumn{1}{c}{$c^{(\bar I)}_{5,3}$} 
\\
\hline
\noalign{\smallskip}
\multicolumn{1}{c}{${\frac {351872}{175175}}$} & 
\multicolumn{1}{c}{$-{\frac {7225}{21021}}$} & 
\multicolumn{1}{c}{$-{\frac {13747
}{64680}}$} &
\multicolumn{1}{c}{$-{\frac {787}{24024}}$} & 
\multicolumn{1}{c}{$-{\frac 
{5}{3003}}$} & 
\multicolumn{1}{c}{${\frac {125632}{40425}}$} & 
\multicolumn{1}{c}{$-{\frac {636578}{1576575}}$} & 
\multicolumn{1}{c}{$-{\frac {
2131}{4550}}$} & 
\multicolumn{1}{c}{$-{\frac {421523}{3603600}}$} \\
\noalign{\smallskip}
\multicolumn{1}{c}{$c^{(\bar I)}_{5,4}$}
& \multicolumn{1}{c}{$c^{(\bar I)}_{5,5}$} 
& \multicolumn{1}{c}{$c^{(\bar I)}_{6,0}$} 
& \multicolumn{1}{c}{$c^{(\bar I)}_{6,1}$} 
& \multicolumn{1}{c}{$c^{(\bar I)}_{6,2}$} 
& \multicolumn{1}{c}{$c^{(\bar I)}_{6,3}$} 
& \multicolumn{1}{c}{$c^{(\bar I)}_{6,4}$} 
& \multicolumn{1}{c}{$c^{(\bar I)}_{6,5}$} 
& \multicolumn{1}{c}{$c^{(\bar I)}_{6,6}$} 
\\
\hline
\noalign{\smallskip}
 \multicolumn{1}{c}{$-{\frac {601}{48048}}
$} & 
 \multicolumn{1}{c}{$-{\frac{1}{1980}}$} & 
 \multicolumn{1}{c}{${\frac {60771136}{12182625}}$} &
 \multicolumn{1}{c}{$-{\frac {9147988}{26801775}}$} & 
 \multicolumn{1}{c}{$-{
\frac {8480897}{8933925}}$} & 
 \multicolumn{1}{c}{$-{\frac {146991919}{428828400}}$} & 
 \multicolumn{1}{c}{$-{\frac {8218907}{142942800}}$} & 
 \multicolumn{1}{c}{$-{
\frac {4207}{875160}}$} & 
 \multicolumn{1}{c}{$-{\frac {71}{
437580}}$} \\
\noalign{\smallskip}
\hline
\hline
%\end{tabular}
%\end{centering}
%\caption{\label{Table:Ibar} Coefficients $c^{(\bar I)}_{ij}$ in Eq.~\eqref{eq:Ibar-PM} for $\bar I$ as a function of the stellar compactness $C$ within the PM approximation.}
%\end{table*}
\endfulltable

%
%{\renewcommand{\arraystretch}{1.2}
%\begin{table*}
%\begin{centering}
%\begin{tabular}{cccccccccccccccccccccccc}
\fulltable{\label{Table:lambdabar} Coefficients $c^{(\bar \lambda_2)}_{ij}$ in Eq.~\eqref{eq:lambdabar-PM} for $\bar \lambda_2$ as a function of the stellar compactness $C$ within the PM approximation.
\vspace{3mm}}
\hline
\hline
\noalign{\smallskip}
\multicolumn{1}{c}{$c^{(\bar \lambda_2)}_{1,0}$}
& \multicolumn{1}{c}{$c^{(\bar \lambda_2)}_{1,1}$} 
& \multicolumn{1}{c}{$c^{(\bar \lambda_2)}_{1,2}$} 
& \multicolumn{1}{c}{$c^{(\bar \lambda_2)}_{2,0}$} 
& \multicolumn{1}{c}{$c^{(\bar \lambda_2)}_{2,1}$} 
& \multicolumn{1}{c}{$c^{(\bar \lambda_2)}_{2,2}$} 
\\
\hline
\noalign{\smallskip}
\multicolumn{1}{c}{$-{\frac {160}{7}}$} & 
\multicolumn{1}{c}{$-{\frac {995}{42}}$} & 
\multicolumn{1}{c}{${\frac {115}{84}}$} &
\multicolumn{1}{c}{${\frac {80480}{441}}$} & 
\multicolumn{1}{c}{${\frac {461660}{1323}}$} & 
\multicolumn{1}{c}{${\frac {82955}{588}}$} \\
\noalign{\smallskip}
 \multicolumn{1}{c}{$c^{(\bar \lambda_2)}_{2,3}$}
& \multicolumn{1}{c}{$c^{(\bar \lambda_2)}_{2,4}$}
& \multicolumn{1}{c}{$c^{(\bar \lambda_2)}_{3,0}$} 
& \multicolumn{1}{c}{$c^{(\bar \lambda_2)}_{3,1}$} 
& \multicolumn{1}{c}{$c^{(\bar \lambda_2)}_{3,2}$} 
& \multicolumn{1}{c}{$c^{(\bar \lambda_2)}_{3,3}$} 
\\
\hline
\noalign{\smallskip}
\multicolumn{1}{c}{$-{\frac {280}{9}}$} & 
\multicolumn{1}{c}{$-{\frac {2825}{3024}}$} & 
\multicolumn{1}{c}{$-{\frac {6598400}{11319}}$} & 
\multicolumn{1}{c}{$-{\frac {143668582}{101871}}$} & 
\multicolumn{1}{c}{$-{
\frac {2135837}{2058}}$} & 
\multicolumn{1}{c}{$-{\frac {4341409
}{58212}}$} \\
\noalign{\smallskip}
 \multicolumn{1}{c}{$c^{(\bar \lambda_2)}_{3,4}$}
& \multicolumn{1}{c}{$c^{(\bar \lambda_2)}_{3,5}$} 
& \multicolumn{1}{c}{$c^{(\bar \lambda_2)}_{3,6}$} 
& \multicolumn{1}{c}{$c^{(\bar \lambda_2)}_{4,0}$} 
& \multicolumn{1}{c}{$c^{(\bar \lambda_2)}_{4,1}$} 
& \multicolumn{1}{c}{$c^{(\bar \lambda_2)}_{4,2}$}
\\
\hline
\noalign{\smallskip}
\multicolumn{1}{c}{${\frac {58914551}{465696}}$} & 
\multicolumn{1}{c}{$-{\frac {10915}{2464}}$} & 
\multicolumn{1}{c}{$-{\frac {322075}{266112}}$} & 
\multicolumn{1}{c}{${\frac {1713843200}{3090087}}$} & 
\multicolumn{1}{c}{${\frac {6610428752}{9270261}}$} & 
\multicolumn{1}{c}{${\frac {1547606966}{27810783}}$} \\
%\hline
%\hline
\noalign{\smallskip}
 \multicolumn{1}{c}{$c^{(\bar \lambda_2)}_{4,3}$}
& \multicolumn{1}{c}{$c^{(\bar \lambda_2)}_{4,4}$} 
& \multicolumn{1}{c}{$c^{(\bar \lambda_2)}_{4,5}$} 
& \multicolumn{1}{c}{$c^{(\bar \lambda_2)}_{4,6}$} 
& \multicolumn{1}{c}{$c^{(\bar \lambda_2)}_{4,7}$} 
& \multicolumn{1}{c}{$c^{(\bar \lambda_2)}_{4,8}$} \\
\hline
\noalign{\smallskip}
\multicolumn{1}{c}{$-{\frac {351398725}{7945938}}$} & 
\multicolumn{1}{c}{${\frac {234321937}{1629936}}$} &
\multicolumn{1}{c}{${\frac {2331123479}{127135008}}$} & 
\multicolumn{1}{c}{$-{\frac {1030100675}{18162144}}$} & 
\multicolumn{1}{c}{$-{\frac {5990225}{314496}}$} & 
\multicolumn{1}{c}{$-{
\frac {1097875}{628992}}$} \\
\noalign{\smallskip}
 \multicolumn{1}{c}{$c^{(\bar \lambda_2)}_{5,0}$}
& \multicolumn{1}{c}{$c^{(\bar \lambda_2)}_{5,1}$} 
& \multicolumn{1}{c}{$c^{(\bar \lambda_2)}_{5,2}$} 
& \multicolumn{1}{c}{$c^{(\bar \lambda_2)}_{5,3}$} 
& \multicolumn{1}{c}{$c^{(\bar \lambda_2)}_{5,4}$} 
& \multicolumn{1}{c}{$c^{(\bar \lambda_2)}_{5,5}$} \\
\hline
\noalign{\smallskip}
\multicolumn{1}{c}{${\frac {8860516352}{64891827}}$} & 
\multicolumn{1}{c}{${\frac {2217813273272}{
973377405}}$} &
\multicolumn{1}{c}{${\frac {1550789734838}{584026443}}$} & 
\multicolumn{1}{c}{${\frac 
{16971435941}{30900870}}$} & 
\multicolumn{1}{c}{${\frac {21489165403}{31783752}}$} & 
\multicolumn{1}{c}{${\frac {10043710866671}{13349175840
}}$} \\
\noalign{\smallskip}
 \multicolumn{1}{c}{$c^{(\bar \lambda_2)}_{5,6}$}
& \multicolumn{1}{c}{$c^{(\bar \lambda_2)}_{5,7}$} 
& \multicolumn{1}{c}{$c^{(\bar \lambda_2)}_{5,8}$} 
& \multicolumn{1}{c}{$c^{(\bar \lambda_2)}_{5,9}$} 
& \multicolumn{1}{c}{$c^{(\bar \lambda_2)}_{5,10}$} 
& \multicolumn{1}{c}{$c^{(\bar \lambda_2)}_{6,0}$} \\
\hline
\noalign{\smallskip}
\multicolumn{1}{c}{$-{\frac {1749233951}{3302208}}$} & 
\multicolumn{1}{c}{$-{\frac {200852202065
}{254270016}}$} &
\multicolumn{1}{c}{$-{\frac {130013401675}{
435891456}}$} & 
\multicolumn{1}{c}{$-{\frac {13797972625}{290594304}}$} & 
\multicolumn{1}{c}{$-{\frac {
113935625}{41513472}}$} & 
\multicolumn{1}{c}{${\frac {3018408755200}{9438155727}}$} \\
\noalign{\smallskip}
 \multicolumn{1}{c}{$c^{(\bar \lambda_2)}_{6,1}$}
& \multicolumn{1}{c}{$c^{(\bar \lambda_2)}_{6,2}$} 
& \multicolumn{1}{c}{$c^{(\bar \lambda_2)}_{6,3}$} 
& \multicolumn{1}{c}{$c^{(\bar \lambda_2)}_{6,4}$} 
& \multicolumn{1}{c}{$c^{(\bar \lambda_2)}_{6,5}$} 
& \multicolumn{1}{c}{$c^{(\bar \lambda_2)}_{6,6}$} \\
\hline
\noalign{\smallskip}
\multicolumn{1}{c}{${\frac {3742297951628752
}{1274151023145}}$} & 
${\frac {55197038324995828}{
3822453069435}}$ &
\multicolumn{1}{c}{${\frac {472845689999591}{26003082105}}$} & 
${\frac {24915470568822391}{2184258896820}}$ & 
${\frac {
874072035313769}{107864636880}}$ & 
$-{\frac {
958041289000963}{499259176416}}$ \\
\noalign{\smallskip}
 \multicolumn{1}{c}{$c^{(\bar \lambda_2)}_{6,7}$}
& \multicolumn{1}{c}{$c^{(\bar \lambda_2)}_{6,8}$} 
& \multicolumn{1}{c}{$c^{(\bar \lambda_2)}_{6,9}$} 
& \multicolumn{1}{c}{$c^{(\bar \lambda_2)}_{6,10}$} 
& \multicolumn{1}{c}{$c^{(\bar \lambda_2)}_{6,11}$} 
& \multicolumn{1}{c}{$c^{(\bar \lambda_2)}_{6,12}$} \\
\hline
\noalign{\smallskip}
$-{\frac {
88927876896912359}{6656789018880}}$ & 
$-{\frac {
2451739592866309}{207484333056}}$ &
$-{\frac {
1661453222619755}{351127332864}}$ & 
$-{\frac {
698286437561675}{702254665728}}$ &  
$-{\frac {763587990875
}{7185604608}}$ & 
\multicolumn{1}{c}{$-{\frac {568540625}{
124084224}}$} \\
\noalign{\smallskip}
\hline
\hline
%\end{tabular}
%\end{centering}
%\caption{\label{Table:lambdabar} Coefficients $c^{(\bar \lambda_2)}_{ij}$ in Eq.~\eqref{eq:lambdabar-PM} for $\bar \lambda_2$ as a function of the stellar compactness $C$ within the PM approximation.}
%\end{table*}
%}
\endfulltable

%
%{\renewcommand{\arraystretch}{1.2}
%\begin{table*}
%\begin{centering}
%\begin{tabular}{cccccccccccccccccccccccc}
\fulltable{\label{Table:ILove} Coefficients $c^{(\bar I  \bar \lambda_2)}_{ij}$ in Eq.~\eqref{eq:ILove-PM} for $\bar I$ as a function of the stellar compactness $\bar \lambda_2$ within the PM approximation.
\vspace{3mm}}
\hline
\hline
\noalign{\smallskip}
\multicolumn{1}{c}{$c^{(\bar I \bar \lambda_2)}_{1,0}$}
& \multicolumn{1}{c}{$c^{(\bar I \bar \lambda_2)}_{1,1}$} 
& \multicolumn{1}{c}{$c^{(\bar I \bar \lambda_2)}_{1,2}$} 
& \multicolumn{1}{c}{$c^{(\bar I \bar \lambda_2)}_{2,0}$} 
& \multicolumn{1}{c}{$c^{(\bar I \bar \lambda_2)}_{2,1}$} 
\\
\hline
\noalign{\smallskip}
\multicolumn{1}{c}{$\frac{88}{7}$} & 
\multicolumn{1}{c}{$\frac{40}{3}$} & 
\multicolumn{1}{c}{$-\frac{13}{12}$} &
\multicolumn{1}{c}{$\frac {139616}{2205}$} & 
\multicolumn{1}{c}{$\frac {196360}{1323}$}  
\\
\noalign{\smallskip}
 \multicolumn{1}{c}{$c^{(\bar I \bar \lambda_2)}_{2,2}$} 
& \multicolumn{1}{c}{$c^{(\bar I \bar \lambda_2)}_{2,3}$}
& \multicolumn{1}{c}{$c^{(\bar I \bar \lambda_2)}_{2,4}$}
& \multicolumn{1}{c}{$c^{(\bar I \bar \lambda_2)}_{3,0}$} 
& \multicolumn{1}{c}{$c^{(\bar I \bar \lambda_2)}_{3,1}$} 
\\
\hline
\noalign{\smallskip}
\multicolumn{1}{c}{$\frac {19127}{252}$} &
\multicolumn{1}{c}{$-{\frac {11219}{1176}}$} & 
\multicolumn{1}{c}{${\frac {3175}{10584}}$} & 
\multicolumn{1}{c}{${\frac {5109760}{33957}}$} & 
\multicolumn{1}{c}{${\frac {98970884}{169785}}$}  
\\
\noalign{\smallskip}
 \multicolumn{1}{c}{$c^{(\bar I \bar \lambda_2)}_{3,2}$} 
& \multicolumn{1}{c}{$c^{(\bar I \bar \lambda_2)}_{3,3}$} 
& \multicolumn{1}{c}{$c^{(\bar I \bar \lambda_2)}_{3,4}$}
& \multicolumn{1}{c}{$c^{(\bar I \bar \lambda_2)}_{3,5}$} 
& \multicolumn{1}{c}{$c^{(\bar I \bar \lambda_2)}_{3,6}$} 
\\
\hline
\noalign{\smallskip}
\multicolumn{1}{c}{${\frac {1148580899}{1528065}}$} & 
\multicolumn{1}{c}{${\frac {1850685733}{6112260}}$} &
\multicolumn{1}{c}{$-{\frac {30603863}{2037420}}$} & 
\multicolumn{1}{c}{$-{\frac {
4293193}{1222452}}$} & 
\multicolumn{1}{c}{$-{\frac {10717951}{
9779616}}$} 
\\
%\hline
%\hline
\noalign{\smallskip}
 \multicolumn{1}{c}{$c^{(\bar I \bar \lambda_2)}_{4,0}$} 
& \multicolumn{1}{c}{$c^{(\bar I \bar \lambda_2)}_{4,1}$} 
& \multicolumn{1}{c}{$c^{(\bar I \bar \lambda_2)}_{4,2}$}
& \multicolumn{1}{c}{$c^{(\bar I \bar \lambda_2)}_{4,3}$}
& \multicolumn{1}{c}{$c^{(\bar I \bar \lambda_2)}_{4,4}$} 
\\
\hline
\noalign{\smallskip}
\multicolumn{1}{c}{${\frac {2719077376}{21068775}}$} & 
\multicolumn{1}{c}{${\frac {448894288}{567567}}$} & 
\multicolumn{1}{c}{${\frac {73208568884}{37923795}}$} &
\multicolumn{1}{c}{${\frac {129979163705}{55621566}}$} & 
\multicolumn{1}{c}{${\frac {70155024767}{
57047760}}$} 
\\
\noalign{\smallskip}
 \multicolumn{1}{c}{$c^{(\bar I \bar \lambda_2)}_{4,5}$} 
& \multicolumn{1}{c}{$c^{(\bar I \bar \lambda_2)}_{4,6}$} 
& \multicolumn{1}{c}{$c^{(\bar I \bar \lambda_2)}_{4,7}$} 
& \multicolumn{1}{c}{$c^{(\bar I \bar \lambda_2)}_{4,8}$} 
& \multicolumn{1}{c}{$c^{(\bar I \bar \lambda_2)}_{5,0}$}
\\
\hline
\noalign{\smallskip}
\multicolumn{1}{c}{${\frac {1602346982741}{13349175840}}$} & 
\multicolumn{1}{c}{$-{\frac {1285405448827}{17798901120}}$} & 
\multicolumn{1}{c}{$-{\frac {13875033761}{889945056}}$} & 
\multicolumn{1}{c}{$-{\frac {15389455007}{42717362688}}$} &
\multicolumn{1}{c}{$-{\frac {507262148608}{4866887025}}$} 
\\
\noalign{\smallskip}
 \multicolumn{1}{c}{$c^{(\bar I \bar \lambda_2)}_{5,1}$} 
& \multicolumn{1}{c}{$c^{(\bar I \bar \lambda_2)}_{5,2}$} 
& \multicolumn{1}{c}{$c^{(\bar I \bar \lambda_2)}_{5,3}$} 
& \multicolumn{1}{c}{$c^{(\bar I \bar \lambda_2)}_{5,4}$} 
& \multicolumn{1}{c}{$c^{(\bar I \bar \lambda_2)}_{5,5}$} 
\\
\hline
\noalign{\smallskip}
\multicolumn{1}{c}{$-{\frac {7677705691856}{14600661075}}$} &
\multicolumn{1}{c}{$-{\frac {2244933136724}{3369383325}}$} & 
\multicolumn{1}{c}{${\frac {1619202392194}{1706570775}}$} & 
\multicolumn{1}{c}{$ {\frac {212537186204347}{52562379870}}$} & 
\multicolumn{1}{c}{${\frac {24258802978997}{5990014800}}$} 
\\
\noalign{\smallskip}
 \multicolumn{1}{c}{$c^{(\bar I \bar \lambda_2)}_{5,6}$}
& \multicolumn{1}{c}{$c^{(\bar I \bar \lambda_2)}_{5,7}$} 
& \multicolumn{1}{c}{$c^{(\bar I \bar \lambda_2)}_{5,8}$} 
& \multicolumn{1}{c}{$c^{(\bar I \bar \lambda_2)}_{5,9}$} 
& \multicolumn{1}{c}{$c^{(\bar I \bar \lambda_2)}_{5,10}$} 
\\
\hline
\noalign{\smallskip}
\multicolumn{1}{c}{${\frac {1882420967949367}{1681996155840}}$} & 
\multicolumn{1}{c}{$-{\frac {28481247349223}{120142582560}}$} &
\multicolumn{1}{c}{$-{\frac {39174857283707}{320380220160}}$} & 
\multicolumn{1}{c}{$-{\frac {1912091255831}{96114066048}}$} & 
\multicolumn{1}{c}{$ -{\frac {2754748757447}{1922281320960}}$} 
\\
\noalign{\smallskip}
 \multicolumn{1}{c}{$c^{(\bar I \bar \lambda_2)}_{6,0}$} 
& \multicolumn{1}{c}{$c^{(\bar I \bar \lambda_2)}_{6,1}$}
& \multicolumn{1}{c}{$c^{(\bar I \bar \lambda_2)}_{6,2}$} 
& \multicolumn{1}{c}{$c^{(\bar I \bar \lambda_2)}_{6,3}$} 
& \multicolumn{1}{c}{$c^{(\bar I \bar \lambda_2)}_{6,4}$} 
\\
\hline
\noalign{\smallskip}
$-{\frac {120917752852250624}{286683980207625}}$ &
$-{\frac {214334584423866976}{57336796041525}}$ & 
$-{\frac {2400708364625585884}{172010388124575}}$ &
$-{\frac {43101354764736391546}{1548093493121175}}$ & 
$-{\frac {22084209085559827543}{688041552498300}}$ 
\\
\noalign{\smallskip}
 \multicolumn{1}{c}{$c^{(\bar I \bar \lambda_2)}_{6,5}$} 
& \multicolumn{1}{c}{$c^{(\bar I \bar \lambda_2)}_{6,6}$} 
& \multicolumn{1}{c}{$c^{(\bar I \bar \lambda_2)}_{6,7}$}
& \multicolumn{1}{c}{$c^{(\bar I \bar \lambda_2)}_{6,8}$} 
& \multicolumn{1}{c}{$c^{(\bar I \bar \lambda_2)}_{6,9}$} 
\\
\hline
\noalign{\smallskip}
$ -{\frac {249310930642237571}{19112265347175}}$ & 
${\frac {333660111817466369461}{33025994519918400}}$ &
$\frac{11506518464513510959}{1000787712724800}$ & 
$\frac{8299909121934464119}{2795851387929600}$ &
$-\frac{2919998240988289177}{9607562042158080}$  
\\
\noalign{\smallskip}
 \multicolumn{1}{c}{$c^{(\bar I \bar \lambda_2)}_{6,10}$} 
& \multicolumn{1}{c}{$c^{(\bar I \bar \lambda_2)}_{6,11}$} 
& \multicolumn{1}{c}{$c^{(\bar I \bar \lambda_2)}_{6,12}$} 
&
&
\\
\hline
\noalign{\smallskip}
$-\frac{8676262683332995337}{35227727487912960}$ &  
$-\frac{1036509722405634683}{28182181990330368}$ & 
$-\frac{8375090116034335069}{5072792758259466240}$ &
&
\\
\noalign{\smallskip}
\hline
\hline
%\end{tabular}
%\end{centering}
%\caption{\label{Table:ILove} Coefficients $c^{(\bar I  \bar \lambda_2)}_{ij}$ in Eq.~\eqref{eq:ILove-PM} for $\bar I$ as a function of the stellar compactness $\bar \lambda_2$ within the PM approximation. }
%\end{table*}
%}
\endfulltable

In this appendix, we show some of the coefficients in the PM expressions of Sec.~\ref{sec:PM}, and explain how one can construct Pad\'e approximants following~\cite{Chan:2014tva}. The constants $c^{(\bar I)}_{ij}$,  $c^{(\bar \lambda_2)}_{ij}$ and $c^{(\bar I  \bar \lambda_2)}_{ij}$ in Eqs.~\eqref{eq:Ibar-PM}--\eqref{eq:ILove-PM} in GR are given in Tables~\ref{Table:Ibar}--\ref{Table:ILove} respectively.
Notice that Eqs.~\eqref{eq:Ibar-PM}--\eqref{eq:ILove-PM} all have the form
\be
\label{eq:6th}
y = \alpha_0 x^k \sum_{i=1}^6 \left[ 1 + \alpha_i x^i + \mathcal{O}\left( x^7 \right)  \right]\,,
\ee
where $x=C$ for Eqs.~\eqref{eq:Ibar-PM} and~\eqref{eq:lambdabar-PM}, while $x=\bar \lambda_2^{-1/5}$ for Eq.~\eqref{eq:ILove-PM}. From Eq.~\eqref{eq:6th}, one can construct the (3,3)-Pad\'e approximant given by
\be
\label{eq:33Pade}
y = \alpha_0 x^k \left[ \frac{\sum_{i=0}^3 \beta_{1i}^{(3)} x^i}{\sum_{i=0}^3 \beta_{2i}^{(3)} x^i} + \mathcal{O}\left( x^7 \right) \right]\,,
\ee
with the following identification of constants
\allowdisplaybreaks
%\bw
\begin{align}
\beta_{10}^{(3)} &= \alpha_1\,\alpha_3\,\alpha_5-\alpha_1\,{\alpha_4}^{2}-{\alpha_2}^{2}\alpha_5+
2\,\alpha_2\,\alpha_3\,\alpha_4-{\alpha_3}^{3}\,, \\
\beta_{11}^{(3)} &= \left( \alpha_3\,\alpha_5-{\alpha_4}^{2} \right) {\alpha_1}^{2}+ \left[ -
{\alpha_3}^{3} + \left( 2\,\alpha_2\,\alpha_4-\alpha_6 \right) \alpha_3-
\alpha_5\, \left( {\alpha_2}^{2}-\alpha_4 \right)  \right] \alpha_1 \nn \\
&+{
\alpha_2}^{2}\alpha_6-\alpha_2\,\alpha_3\,\alpha_5-\alpha_2\,{\alpha_4}^{2}+{
\alpha_3}^{2}\alpha_4\,,  \\
\beta_{12}^{(3)} &= \left( -\alpha_3\,\alpha_6+\alpha_4\,\alpha_5 \right) {\alpha_1}^{2}+
 \left( {\alpha_2}^{2}\alpha_6-2\,\alpha_2\,{\alpha_4}^{2}+{\alpha_3}^{2}
\alpha_4+\alpha_4\,\alpha_6-{\alpha_5}^{2} \right) \alpha_1 \nn \\
&-{\alpha_2}^{3}
\alpha_5+2\,{\alpha_2}^{2}\alpha_3\,\alpha_4 +\left( -{\alpha_3}^{3}-\alpha_3\,\alpha_6+\alpha_4\,\alpha_5 \right) \alpha_2+{\alpha_3}^{2}\alpha_5-
\alpha_3\,{\alpha_4}^{2}\,, \\
\beta_{13}^{(3)} &= \alpha_6\,{\alpha_2}^{3}+ \left( -2\,\alpha_3\,\alpha_5-{\alpha_4}^{2}
 \right) {\alpha_2}^{2}+ \left[ 3\,{\alpha_3}^{2}\alpha_4-2\,\alpha_1\,
\alpha_3\,\alpha_6+ \left( 2\,\alpha_1\,\alpha_5-\alpha_6 \right) \alpha_4 \right. \nn \\
& \left. +{
\alpha_5}^{2} \right] \alpha_2-{\alpha_3}^{4} + \left( 2\,\alpha_1\,\alpha_5
+\alpha_6 \right) {\alpha_3}^{2} + \left( -2\,\alpha_1\,{\alpha_4}^{2}-2\,
\alpha_4\,\alpha_5 \right) \alpha_3 \nn \\
& +{\alpha_1}^{2}\alpha_4\,\alpha_6-{\alpha_1}^{2}{\alpha_5}^{2}+{\alpha_4}^{3}\,, \\
\beta_{20}^{(3)} &=\beta_{10}^{(3)}\,, \\
\beta_{21}^{(3)} &= -\alpha_2\,{\alpha_4}^{2}+ \left( \alpha_1\,\alpha_5+{\alpha_3}^{2}
 \right) \alpha_4+ \left( -\alpha_1\,\alpha_6-\alpha_2\,\alpha_5 \right) 
\alpha_3+{\alpha_2}^{2}\alpha_6\,, \\
\beta_{22}^{(3)} &= -\alpha_1\,{\alpha_5}^{2}+ \left( \alpha_2\,\alpha_4+{\alpha_3}^{2}
 \right) \alpha_5+\alpha_1\,\alpha_4\,\alpha_6-\alpha_2\,\alpha_3\,\alpha_6-
\alpha_3\,{\alpha_4}^{2}\,, \\
\beta_{23}^{(3)} &= -\alpha_2\,\alpha_4\,\alpha_6+\alpha_2\,{\alpha_5}^{2}+{\alpha_3}^{2}\alpha_6
-2\,\alpha_3\,\alpha_4\,\alpha_5+{\alpha_4}^{3}\,. 
\end{align}
%\ew
%

In dCS gravity, we derived PM relations between the dimensionless dipole scalar charge and the compactness, given as
\be
\label{eq:mubar-PM}
\bar \mu =  \frac{64}{5} \sum_{i=1}^{4} \sum_{j=1}^{i} \left[ 1 + c_{ij}^{(\bar \mu)}  \left(\frac{\lambda_\BL}{7 \pi} \right)^j C^i + \mathcal{O} \left( C^5 \right)  \right]\,,
\ee
where coefficients $c_{ij}^{(\bar \mu)}$ are given in Table~\ref{Table:mubar}. Equation~\eqref{eq:mubar-PM} has the form of Eq.~\eqref{eq:6th} to $\mathcal{O}(x^5)$. As we did for Eq.~\eqref{eq:33Pade}, one can then construct the (2,2)-Pad\'e approximant 
\be
\label{eq:22Pade}
y = \alpha_0 x^k \left[ \frac{\sum_{i=0}^2 \beta_{1i}^{(2)} x^i}{\sum_{i=0}^3 \beta_{2i}^{(2)} x^i} + \mathcal{O}\left( x^5 \right) \right]\,,
\ee
by identifying the constants
%
%\allowdisplaybreaks
%\bw
\begin{align}
\beta_{10}^{(2)} &= \alpha_1\,\alpha_3-{\alpha_2}^{2}\,, \\
\beta_{11}^{(2)} &= {\alpha_1}^{2}\alpha_3+ \left( -{\alpha_2}^{2}-\alpha_4 \right) \alpha_1+
\alpha_2\,\alpha_3\,, \\ % \nn \\ 
\beta_{12}^{(2)} &= -{\alpha_2}^{3}+ \left( 2\,\alpha_1\,\alpha_3+\alpha_4 \right) \alpha_2-{
\alpha_1}^{2}\alpha_4-{\alpha_3}^{2}\,, \\
\beta_{20}^{(2)} &= \beta_{10}^{(2)}\,, \\
\beta_{21}^{(2)} &= -\alpha_1\,\alpha_4+\alpha_2\,\alpha_3\,, \\
\beta_{22}^{(2)} &= \alpha_2\,\alpha_4-{\alpha_3}^{2}\,. 
\end{align}
%

%
%{\renewcommand{\arraystretch}{1.2}
%\begin{table}
%\begin{centering}
%\begin{tabular}{ccccccccccccccccccccccccccc}
\fulltable{\label{Table:mubar} Coefficients $c^{(\bar \mu)}_{ij}$ in Eq.~\eqref{eq:mubar-PM} for $\bar \mu$ as a function of the stellar compactness $C$ within the PM approximation.
\vspace{3mm}}
\hline
\hline
\noalign{\smallskip}
\multicolumn{1}{c}{$c^{(\bar I)}_{1,0}$}
& \multicolumn{1}{c}{$c^{(\bar I)}_{1,1}$} 
& \multicolumn{1}{c}{$c^{(\bar I)}_{2,0}$} 
& \multicolumn{1}{c}{$c^{(\bar I)}_{2,1}$} 
& \multicolumn{1}{c}{$c^{(\bar I)}_{2,2}$} 
& \multicolumn{1}{c}{$c^{(\bar I)}_{3,0}$} 
& \multicolumn{1}{c}{$c^{(\bar I)}_{3,1}$} 
%& \multicolumn{1}{c}{$c^{(\bar I)}_{3,2}$} 
%& \multicolumn{1}{c}{$c^{(\bar I)}_{3,3}$} 
\\
\hline
\noalign{\smallskip}
\multicolumn{1}{c}{$-\frac{3}{4}$} & 
\multicolumn{1}{c}{${\frac {3}{4}}$} & 
\multicolumn{1}{c}{$-{\frac {2}{35}}$} &
\multicolumn{1}{c}{${\frac {34}{105}}$} & 
\multicolumn{1}{c}{$\frac{19}{12}$} & 
\multicolumn{1}{c}{$-{\frac {76}{2205}}$} & 
\multicolumn{1}{c}{${\frac {1481}{2205}}$} & 
%$-{\frac {155}{1848}}$ & 
%$-{\frac {1}{168}}$ 
\\
\noalign{\smallskip}
\multicolumn{1}{c}{$c^{(\bar I)}_{3,2}$}
& \multicolumn{1}{c}{$c^{(\bar I)}_{3,3}$} 
& \multicolumn{1}{c}{$c^{(\bar I)}_{4,0}$} 
& \multicolumn{1}{c}{$c^{(\bar I)}_{4,1}$} 
& \multicolumn{1}{c}{$c^{(\bar I)}_{4,2}$} 
& \multicolumn{1}{c}{$c^{(\bar I)}_{4,3}$} 
& \multicolumn{1}{c}{$c^{(\bar I)}_{4,4}$} 
%& \multicolumn{1}{c}{$c^{(\bar I)}_{5,2}$} 
%& \multicolumn{1}{c}{$c^{(\bar I)}_{5,3}$} 
\\
\hline
\noalign{\smallskip}
\multicolumn{1}{c}{${\frac {186031}{55440}}$} & 
\multicolumn{1}{c}{${\frac {1873}{528}}$} & 
\multicolumn{1}{c}{$-{\frac {1451}{121275}}$} &
\multicolumn{1}{c}{${\frac {115371827}{88288200}}$} & 
\multicolumn{1}{c}{${\frac{38537329}{5045040}}$} & 
\multicolumn{1}{c}{${\frac {40345817}{2882880}}$} & 
\multicolumn{1}{c}{${\frac {686755}{82368}}$} & 
%$-{\frac {2131}{4550}}$ & 
%$-{\frac {421523}{3603600}}$
 \\
\noalign{\smallskip}
\hline
\hline
%\end{tabular}
%\end{centering}
%\caption{\label{Table:mubar} Coefficients $c^{(\bar \mu)}_{ij}$ in Eq.~\eqref{eq:mubar-PM} for $\bar \mu$ as a function of the stellar compactness $C$ within the PM approximation.}
%\end{table}
%}
\endfulltable

%%%%%%%%%%%%%%%%%%%%%%%%%%%%%%%%%%%%%%%%%%%%
%%%%%%%%%%%%%%%%%%%%%%%%%%%%%%%%%%%%%%%%%%%%
\section*{References}

\bibliographystyle{iopart-num}
\bibliography{master}
\end{document}